\documentclass[journal,twoside,final]{IEEEtran}
\usepackage{balance} 
\usepackage{cite}
\usepackage{graphicx}
\usepackage{textcomp}
\def\BibTeX{{\rm B\kern-.05em{\sc i\kern-.025em b}\kern-.08em
    T\kern-.1667em\lower.7ex\hbox{E}\kern-.125emX}}
\usepackage{multirow}
\usepackage{url}
\usepackage{bm}
\usepackage{color}
\usepackage[LY1]{fontenc}
\usepackage{comment}
\usepackage{array}
\usepackage{subcaption}
\usepackage{afterpage}
\newcolumntype{P}[1]{>{\centering\arraybackslash}p{#1}}

\usepackage{amsmath,amssymb,amsfonts}
\usepackage{threeparttable}
\usepackage{siunitx}
\usepackage{soul}
\usepackage{booktabs}

\usepackage{algorithm}
\usepackage{algorithmic}

\DeclareMathSizes{10}{9}{7}{5}

\begin{document}

\title{Real-Time Multi-Level Neonatal Heart and Lung Sound Quality Assessment for Telehealth Applications}
\author{Ethan Grooby, ~\IEEEmembership{Student Member, IEEE}, Chiranjibi Sitaula, Davood Fattahi, Reza Sameni,~\IEEEmembership{Senior Member,~IEEE}, Kenneth Tan, Lindsay Zhou, Arrabella King, Ashwin Ramanathan, Atul Malhotra, Guy A. Dumont, ~\IEEEmembership{Life Fellow,~ IEEE}, Faezeh Marzbanrad, ~\IEEEmembership{Senior Member,~IEEE}
\thanks{E. Grooby, C. Sitaula, D. Fattahi and F. Marzbanrad are with the Department of Electrical and Computer Systems Engineering, Monash University, Melbourne, Australia.}
\thanks{R. Sameni is with the Department of Biomedical Informatics, Emory University, Atlanta, USA}
\thanks{K. Tan, L. Zhou, A. King, A. Ramanathan and A. Malhotra are with Monash Newborn, Monash Children’s
Hospital and Department of Paediatrics, Monash University, Melbourne, Australia.}
\thanks{E. Grooby and G. Dumont are with the Electrical and Computer Engineering Department, University of British Columbia, Vancouver, Canada}
\thanks{E. Grooby  acknowledges the support of MIME-Monash Partners-CSIRO sponsored PhD research support program and Research Training Program (RTP). A. Malhotra research is supported by the Kathleen Tinsley Trust and a Cerebral Palsy Alliance Research Grant. F. Marzbanrad acknowledges the support of Advancing Women's Research Success Grant program. The study is supported by Monash Institute of Medical Engineering (MIME).}

\thanks{Corresponding author: E. Grooby (e-mail: ethan.grooby@monash.edu).}
}
\maketitle

\begin{abstract}
Through the usage of digital stethoscopes in combination with telehealth, chest sounds can be easily collected and transmitted for remote monitoring and diagnosis. Chest sounds contain important information about a newborn's cardio-respiratory health. However, low-quality recordings complicate the remote monitoring and diagnosis. 
 In this study, a new method is proposed to objectively and automatically assess heart and lung signal quality on a 5-level scale in real-time, and to assess the effect of signal quality on vital sign estimation. 
 For the evaluation, a total of 207 10\,s long chest sounds were taken from 119 preterm and full-term babies. Thirty of the recordings from ten subjects were obtained with synchronous vital signs from the Neonatal Intensive Care Unit (NICU) based on electrocardiogram recordings. As reference, seven annotators independently assessed the signal quality. For automatic quality classification, 400 features were extracted from the chest sounds. After feature selection using minimum redundancy and maximum relevancy algorithm, class balancing, and hyper-parameter optimization, a variety of multi-class and ordinal classification and regression algorithms were trained. Then, heart rate and breathing rate were automatically estimated from the chest sounds using adapted pre-existing methods. 
The results of subject-wise leave-one-out cross-validation show that the best-performing models had a mean squared error (MSE) of 0.487 and 0.612, and balanced accuracy of 56.8\% and 51.2\% for heart and lung qualities, respectively. The best-performing models for real-time analysis (<200\,ms) had MSE of 0.459 and 0.673, and balanced accuracy of 56.7\% and 46.3\%, respectively. 
Our experimental results underscore that increasing the signal quality leads to a reduction in vital sign error, with only high-quality recordings having mean absolute error of less than 5 beats per minute, as required for clinical usage. 
\end{abstract}

\begin{IEEEkeywords}
		Breath sound, heart rate, heart sound, neonatal monitoring, ordinal regression, phonocardiogram (PCG), signal quality assessment, respiration rate, telehealth.
\end{IEEEkeywords}

\maketitle

\section{Introduction}
The neonatal period is the most vulnerable time for survival, with 1.7\% of live births resulting in mortality, totalling 2.4 million worldwide, in 2019 alone \cite{Neonatal19:online}. To address this major issue, the United Nations created the 3.2.2 Sustainable Development Goal, with the aim of reducing neonatal mortality to 1.2\% of live births by 2030 \cite{Goal3Dep88:online}. 
Timely assessment for signs of serious health issues, in particular cardiovascular and respiratory health risks potentially improves neonatal survival and reduces long-term morbidity.  

Since stethoscope-recorded chest sounds contain affluent information about neonatal health status, the usage of telehealth-based digital stethoscopes enables accessible timely assessment in both hospital and home environments \cite{ramanathan2019digital,kevat2017systematic,king2020tools}.
It is, however, limited by low-quality chest sounds, due to the noise from either external environment, other internal body sounds, or the device itself. Low-quality recordings complicate monitoring and diagnosis, or at worse lead to misdiagnosis \cite{grooby2020neonatal,lahav2015questionable}.
Whilst having low-quality chest sounds is unavoidable, identification and exclusion of low-quality recordings helps to improve remote monitoring. Current commercial digital stethoscopes either do not support concurrent listening, preventing real-time quality feedback, or provide live auscultation, making it difficult for non-experts and untrained users to interpret and assess the quality \cite{ramanathan2019digital}. Real-time automated quality assessment of heart and lung sounds would address this gap by assisting the users in obtaining better diagnostic-quality recordings and ensuring the reliability of diagnosis.

Previous research on heart signal quality analysis has mainly focused on binary classification of heart sounds into high and low-quality on adult populations. In our past work, these methods were reviewed in detail, adapted and expanded for neonatal population \cite{grooby2020neonatal}. 
To summarise, heart sound recordings were represented in several ways: time and frequency domain, autocorrelation signal, wavelet decomposed signal and segmented heart signal into S1 and S2 sounds \cite{springer2016automated,shi2019automatic,valderrama2018improving,grzegorczyk2016pcg,akram2018analysis}. Features were then extracted from these representations including statistical features (variance, skewness and kurtosis), predictive fitting coefficients, segmentation quality and agreement, Mel-frequency coefficients (MFCC), entropy and power. 
These features were then used to develop a dynamic classifier with 96\% specificity, 81\% sensitivity and 93\% accuracy. These results were shown to be superior than the individual implementation of past heart signal quality estimation techniques \cite{grooby2020neonatal}. 

To date, limited studies have investigated lung sound quality assessment, either relying on an external reference signal or by generating an artificial set of low and high-quality lung sounds \cite{kala2020,emmanouilidou2017computerized}. In our past work, heart sound quality methods were adapted for lung sound quality analysis, with 86\% specificity, 69\% sensitivity, and 82\% accuracy, for binary classification of low vs high-quality \cite{grooby2020neonatal}. 

The key contribution of this research is the automated real-time multi-level quality rating of neonatal chest sounds, which can guide the user during auscultation and sound recording. The development of a real-time system enables objective assessment of signal quality for the user to obtain better diagnostic quality recordings. Another contribution of this research is to extend our previous binary signal quality classification model to a five-level quality scale. This is achieved by introducing new features, providing more detailed signal quality assessment. Through providing a finer scale signal quality assessment, users are able to make more informed decisions on diagnostic quality of the recordings. Unlike previous studies, the processing time to extract features is assessed to determine applicability of heart and lung sound quality assessment in real-time. Since one of the key purposes of acquiring high-quality sounds is to obtain accurate vital sign estimates, the relationship between heart and lung sound quality with heart and breathing rate accuracy is also assessed using gold standard NICU recordings. 

The rest of this paper is organized as follows. Section~\ref{sec:methods} presents details of the proposed signal quality assessment model. Evaluation and results are presented in Section~\ref{sec:results}. Feasibility of real-time analysis and comparison of signal quality with heart and breathing rate error with the model are discussed in Sections~\ref{sec:real} and \ref{sec:sync}, followed by a discussion in Section~\ref{sec:discussion}. Section~\ref{sec:conclusion} concludes the work, with future perspectives.

\section{Methods}
\label{sec:methods}
\subsection{Data Acquisition and Preprocessing}
The study was conducted at Monash Newborn, Monash Children’s Hospital. It was approved by the Monash Health Human Research Ethics Committee (HREA/18/MonH/471). A total of 318, 60\,s recordings from the right anterior chest of preterm and term newborns were obtained using a digital stethoscope \cite{zhou2020acoustic,ramanathan2020assessment}.  Synchronous electrocardiogram and vital signs for reference heart and breathing rate were collected for 32 recordings, as further detailed in Section~\ref{sec:sync}. The chest sounds were low-pass filtered to avoid aliasing and down-sampled to 4\,kHz. Recordings significantly damaged from artifacts making lung and heart sounds impossible to recover were automatically removed using methods presented in our previous work \cite{grooby2020neonatal}. Next, 10\,s segments heart, breathing or both sounds were manually extracted. After excluding the invalid recordings, a total number of 207 signals (119 subjects) remained, 30 (10 subjects) of which had synchronous vital signs. These 30 recordings were held out only for testing the trained models. 

Since the study focuses on both heart and lung sound quality assessment, two pools were generated from the data, for heart and lung. For this purpose, 207 recordings were filtered with a 4\textsuperscript{th}-order Butterworth bandpass filter with passband frequencies 50-250\,Hz and 200-1000\,Hz, in order to separate heart and lung sounds, respectively. This is a commonly used approach to improve signal quality in commercial stethoscopes and clinical studies, with passband frequencies based on the main frequency bands reported in literature for neonatal heart and lung sounds \cite{nersisson2017heart}.

\subsection{Annotation Sets and Quality Annotations}
\begin{figure}[tb]
\centering
\includegraphics[scale=0.34,trim={0.5cm 0.5cm 0.5cm 0.5cm}]{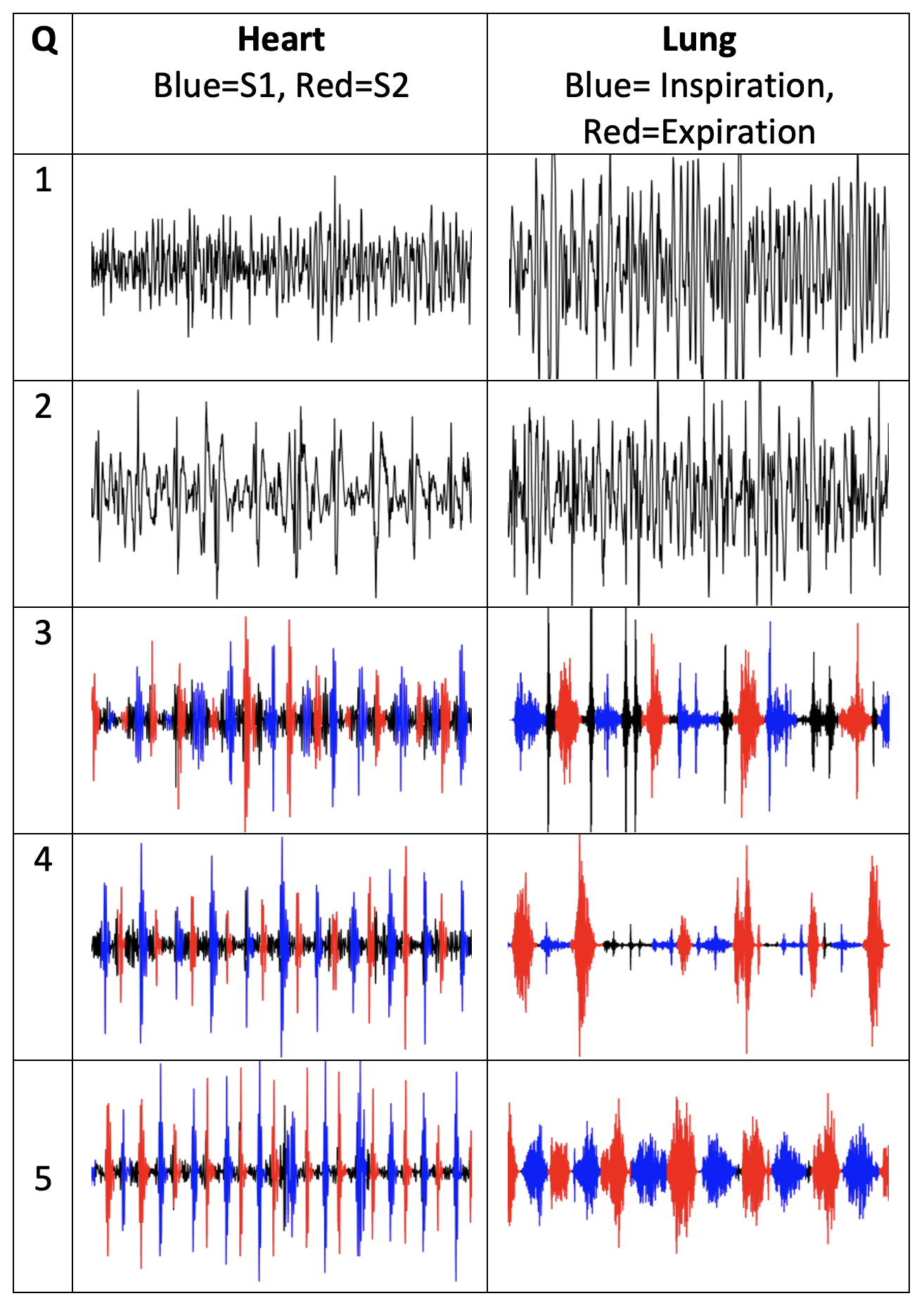}
\caption{Manual annotation guidelines; Q denotes the Signal Quality Label. Example 5\,s plots are shown with heart signal qualities 3-5 segmented into S1 (blue) and S2 (red) using modified method proposed by Springer et al. \cite{springer2015logistic,grooby2020neonatal}, whereas lung signal qualities 3-5 are segmented manually into inspiration (blue) and expiration (red). Signal qualities 1 and 2 were not annotated due to poor quality.}
\label{fig:example_annotation}
\end{figure}

Randomized heart and lung pools were created from 207 raw recordings plus 207 frequency filtered recordings, resulting in 414 heart and 414 lung recording pools. 
Reference annotations for signal quality of recordings were needed in order to develop automatic signal quality estimation method. These were obtained from manually annotated signal quality used as ground truth, provided by 3 clinicians and 4 electrical engineers familiar with biomedical auscultation, producing a set of 5 annotations per recording.
These annotations were performed independently and in a quiet place, by listening to recordings through high-quality earphones. Each recording was assigned an integer score between 1 to 5, with 1 referring to only noise and hardly detectable heart beats/breathing periods, and 5 referring to clear heart/lung sounds with little to no noise. 
Example recordings of signal qualities 1 to 5 are shown in Figure~\ref{fig:example_annotation}.

\begin{figure*}
     \centering
     \begin{subfigure}[b]{0.48\textwidth}
         \centering
         \includegraphics[scale=0.34,trim={0.5cm 0.5cm 0.5cm 0.5cm}]{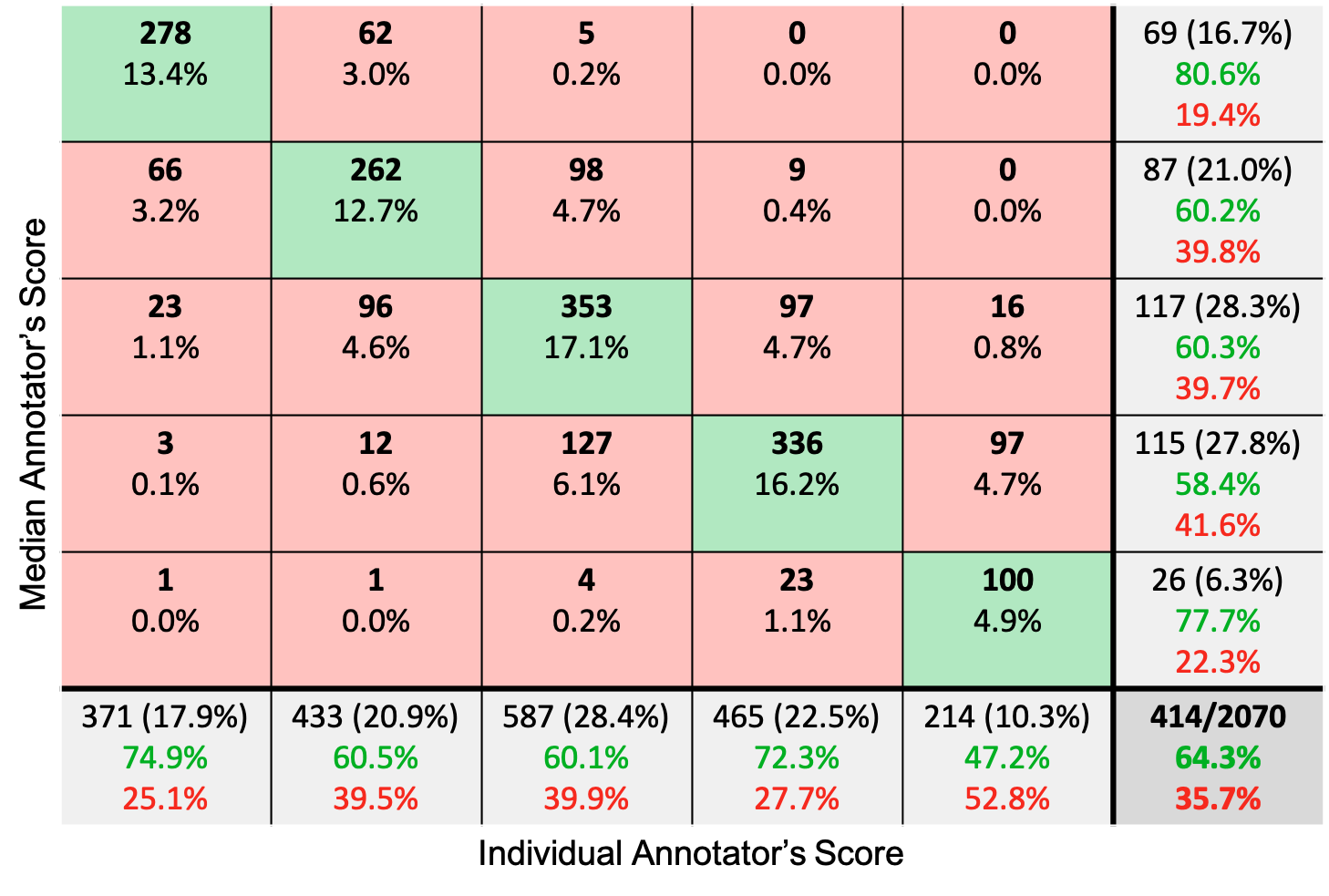}
         \caption{Heart}
     \end{subfigure}
     \begin{subfigure}[b]{0.48\textwidth}
         \centering
         \includegraphics[scale=0.34,trim={0.5cm 0.5cm 0.5cm 0.5cm}]{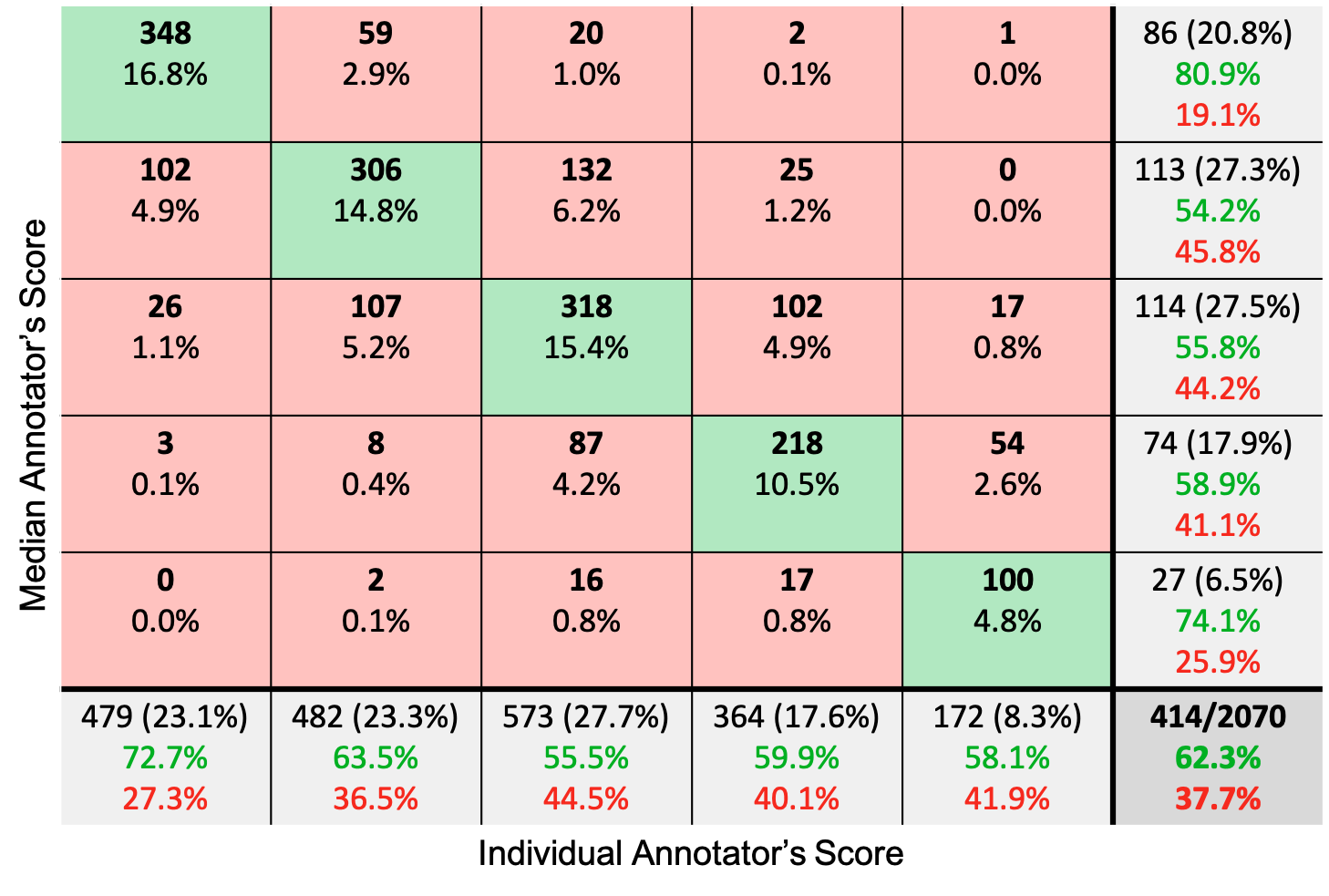}
         \caption{Lung}
     \end{subfigure}
        \caption{Confusion matrix of original annotators' scores}
        \label{fig:HeartLungAnnotationOriginal}
\end{figure*}

\begin{figure*}
     \centering
     \begin{subfigure}[b]{0.48\textwidth}
         \centering
         \includegraphics[scale=0.34,trim={0.5cm 0.5cm 0.5cm 0.5cm}]{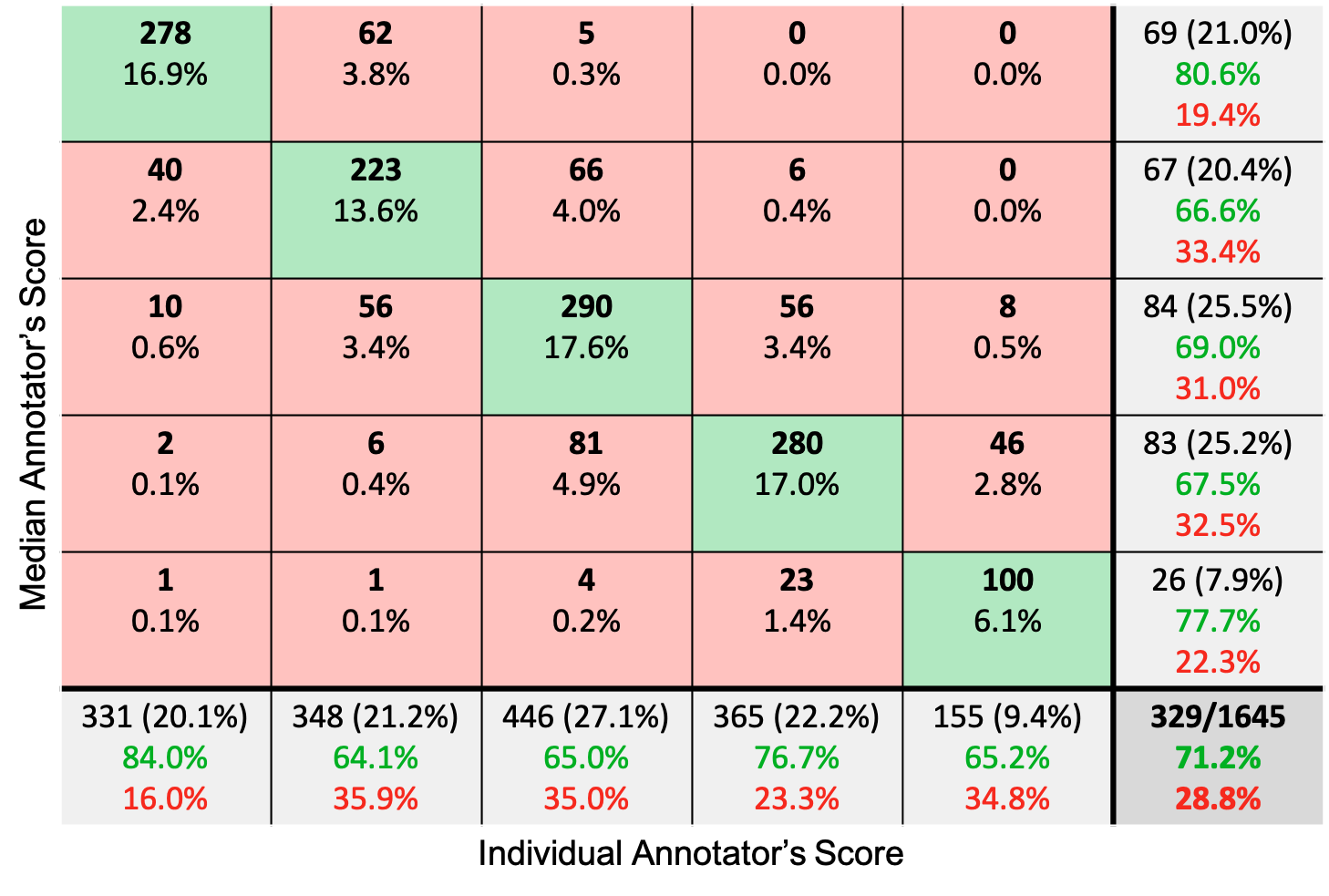}
         \caption{Heart}
     \end{subfigure}
     \begin{subfigure}[b]{0.48\textwidth}
         \centering
         \includegraphics[scale=0.34,trim={0.5cm 0.5cm 0.5cm 0.5cm}]{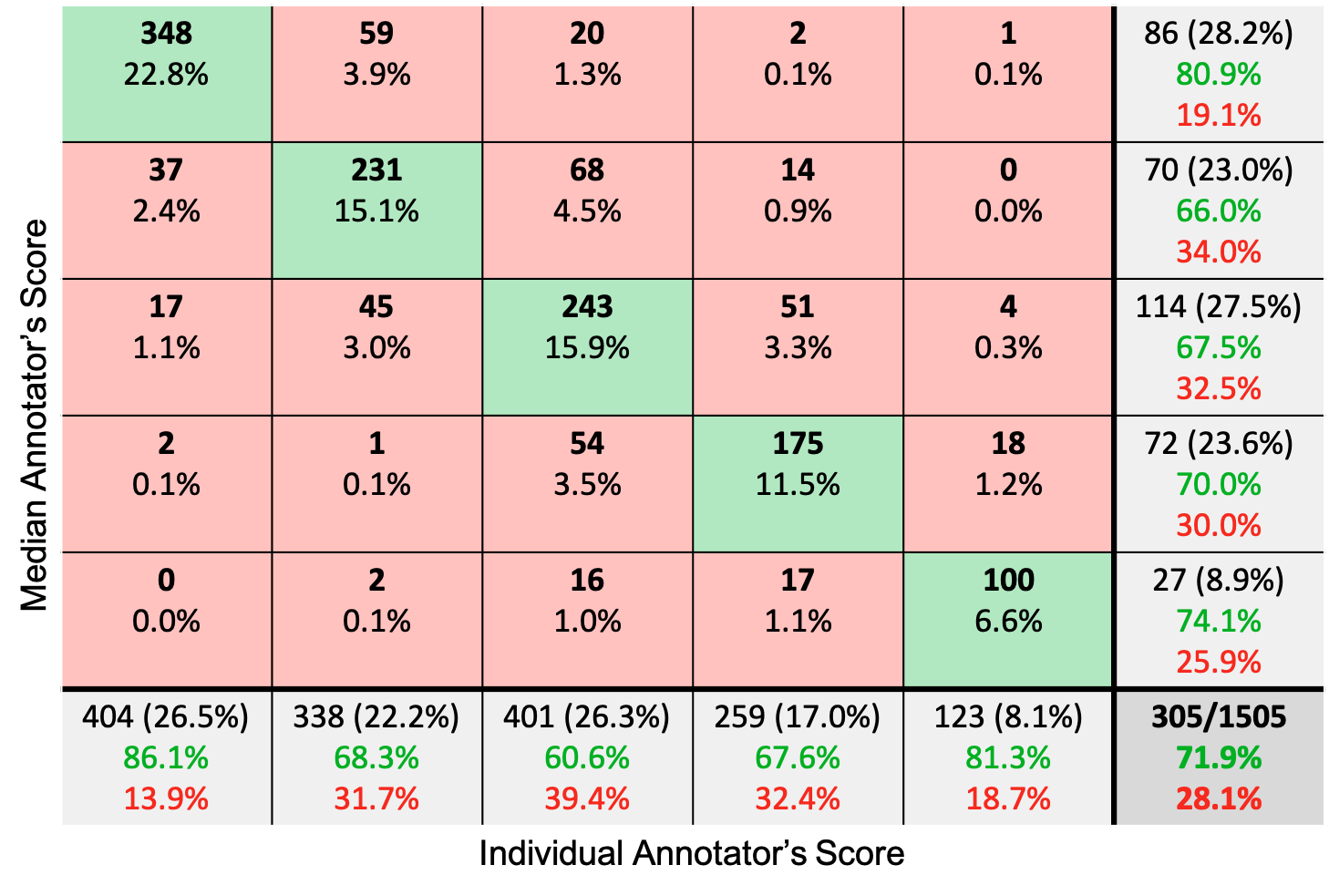}
         \caption{Lung}
     \end{subfigure}
        \caption{Confusion matrix of filtered annotators' scores}
        \label{fig:HeartLungAnnotationFiltered}
\end{figure*}

Inter-rater agreement was evaluated using the Fleiss' kappa \cite{fleiss1971measuring}, with heart and lung agreement scores of 0.28 and 0.27, respectively, which both correspond to fair agreement. 
To have a more reliable training set, recordings with inter-rater agreement less than or equal to 0.2 were removed. This resulted in a total of 329 heart recordings and 305 lung recordings. The short-listed heart and lung records resulted in an inter-rater agreement of 0.37 and 0.39, respectively, which correspond to fair agreement. Median annotators signal quality was then used to determine the signal quality for each recording. The original and resultant signal quality distributions are shown in Figures~\ref{fig:HeartLungAnnotationOriginal} and \ref{fig:HeartLungAnnotationFiltered}, respectively.

\subsection{Quality Assessment Model}
\subsubsection{Features}

\begin{table*}
\caption{Feature set used for automatic classification of heart and lung sounds}
    \centering
    \begin{tabular}{ |p{0.95cm}||p{3cm}|p{12.3cm}| }
    \hline
    \textbf{Number}
    & \textbf{Title}
    & \textbf{Description}
    \\
    \hline
    1
    & Audio Sample Entropy
    & Signal complexity measure which is high for unpredictable signals. Similarity between two epochs determined at a threshold r for M time points is calculated on the down-sampled 30Hz signal with M=2 and r=0.1 \cite{valderrama2018improving}.
    \\
    \hline
    2
    & Percentage Clipping
    & Clipping is an undesirable form of noise due to recorded signal exceeding maximum limit of the digital stethoscope. Normalised magnitude of audio signal (range [0,1]) is calculated and percentage of points above 0.97 is determined \cite{emmanouilidou2017computerized}. 
    \\
    \hline
    3-4
    & Mean Rate Average Energy
    & The average temporal energy variation along each frequency channel of the power spectral density of the recording was calculated over the range 2-32\,Hz \cite{kala2020}. 
    \\
    \hline
    5-6
    & Percentage Heart Contamination
    & Percentage of recording with prominent heart sounds was determined. This was achieved with 50-250\,Hz 4\textsuperscript{th} order Butterworth bandpass filtering of recording. Then wavelet decomposition was performed with Symlet wavelet at depth 3. The 3 approximation coefficients were then normalized and multiplied together to get a representation of prominent heart sound peaks. Percentage of this signal exceeding 0.1 and 0.2 where then calculated \cite{emmanouilidou2017computerized}. 
    \\
    \hline
    7
    & High Frequency Variance
    & Audio signal was 2\textsuperscript{nd} order high pass filtered at 700\,Hz and variance calculated \cite{abdollahpur2017detection,naseri2012computerized}. 
    \\
    \hline
    8-18
    & Linear Predictive Coefficients
    & 10\textsuperscript{th} order linear predictive coefficients were calculated \cite{zabihi2016heart}.
    \\
    \hline
    19-21
    & Entropy
    & Shannon, Renyi and Tsallis entropy of audio recording are calculated \cite{zabihi2016heart}.
    \\
    \hline
    22-23
    & Degree of Periodicity
    & Heart sounds and to an extend breathing sounds can be considered to be quasiperiodic. The the degree in which heart and breathing sounds are periodic in recording are calculated in the ranges 15-220\,bpm and 15-80\,bpm respectively \cite{tang2021automated,li2011best,grooby2020neonatal}.  
    \\
    \hline
    24
    & Autocorrelation Kurtosis
    & Kurtosis of autocorrelation signal is calculated. Kurtosis is a measurement of the degree in which the probability distribution function clusters at the tails \cite{valderrama2018improving}.
    \\
    \hline
    25-26
    & Autocorrelation Sample Entropy
    & Sample entropy with M=2 and r=0.2 is calculated on the 30\,Hz down-sampled full and 5\,s truncated autocorrelation signal \cite{tang2021automated,grooby2020neonatal}.
    \\
    \hline
    27-28
    & Autocorrelation Cycle Duration
    & Cycle duration is calculated based on the peak in the autocorrelation signal corresponding to 70-220\,bpm and 15-80bpm for heart and lung respectively \cite{tang2021automated}.
    \\
    \hline
    29
    & Cry Power
    & Based on previous work, power ratio in 295-406\,Hz is the frequency region in which crying can be most easily identified is calculated \cite{grooby2020neonatal}. 
    \\
    \hline
    30-42
    & Power
    & For 2000\,Hz down-sampled signal, power ratio is determined for 0-100\,Hz, 100-200\,Hz, 200-300\,Hz, 300-400\,Hz, 400-500\,Hz, 500-600\,Hz, 600-700\,Hz, 800-900\,Hz and 900-1000\,Hz frequency ranges \cite{zabihi2016heart}.
    Similarly, this was done for frequency ranges 24-144\,Hz,144-200\,Hz and 200-1000\,Hz \cite{tang2021automated}.
    \\
    \hline
    43
    & Power Centroid
    & For 2000\,Hz down-sampled signal, power spectral density is calculated and centroid calculated \cite{zabihi2016heart}. 
    \\
    \hline
    44-51
    & Linear dependency of PSD
    & The time-frequency power spectral density (PSD) is calculated on original and 2000\,Hz down-sampled signal. The frequency component is next compressed to 15 evenly spaced bins 1-15. Singular value decomposition is then calculated on bins 1-5, 6-10 and 11-15 and ratio of 2\textsuperscript{nd} and 1\textsuperscript{st} components determined\cite{kumar2011noise}. 
    \\
    \hline
    52-61
    & Wavelet Entropy
    & Wavelet decomposition is performed at depth 5 using 4\textsuperscript{th} order Daubechies wavelet. Shannon, Tsallis and Renyi entropies are then calculated on the 5\textsuperscript{th} level approximation coefficients and 3\textsuperscript{rd}, 4\textsuperscript{th} and 5\textsuperscript{th} level detailed coefficients. Additionally log variance is calculated on the 3\textsuperscript{rd} level detailed coefficients \cite{zabihi2016heart}.
    \\
    \hline
    62-63
    & Wavelet RMSSD and ZCR
    & Wavelet decomposition is performed at depth 2 using 8\textsuperscript{th} order Daubechies wavelet. Zero crossing rate (ZCR) and root mean square of successive differences (RMSSD) are calculated on 2\textsuperscript{nd} level approximation coefficients \cite{akram2018analysis}. 
    \\
    \hline
    64-65 
    & Wavelet ZCR
    & Recording was down-sampled to 1000\,Hz and high pass filtered at 20\,Hz. Wavelet decomposition then performed at depth 1 using 2\textsuperscript{nd} order Daubechies wavelet. After normalization, peaks were then detected \cite{gieraltowski2015rs}. ZCR was calculated with threshold of 85\textsuperscript{th} percentile of wavelet decomposed signal and 58\textsuperscript{th} percentile of detected peaks values \cite{grzegorczyk2016pcg}. 
    \\
    \hline
    66-82
    & MFCC
    & MFCC calculated with window length 25ms, overlap length 15ms with 13 coefficients plus log energy. The minimum, maximum and skew is then calculated for all 14 signals and then averaged \cite{zabihi2016heart}. 
    \\
    \hline
    83-84
    & Fundamental Frequency
    & Fundamental frequency calculated using the cepstral method with window length 25ms and overlap 15ms in the 50-1000Hz range. Percentage of frames with fundamental frequency less than 250Hz is calculated as this corresponds to newborn crying \cite{emmanouilidou2017computerized}. Additionally, overall fundamental frequency is determined by plotting the values on a histogram and taking the value of the largest bin \cite{valderrama2018improving}. 
    \\
    \hline
    85-100
    & Envelope Sample Entropy
    & All envelopes were down-sampled to 30Hz and sample entropy with M=2 and r=0.2 was determined \cite{tang2021automated}. 
    \\
    \hline
    101-107
    & Envelope Variance 
    & Variance of all heart based envelopes were calculated \cite{tang2021automated}. 
    \\
    \hline
    108-121
    & Envelope Heart Cycles
    & For heart based envelopes, using the estimated cycle duration from the autocorrelation signal, the envelope is divided up into same length segment and correlation between these segments calculated. The average and standard deviation of the correlation values is then determined \cite{tang2021automated}. 
    \\
    \hline
    122-128
    & Envelope Heart Rate Variability
    & Using a sliding window of 3\,s, heart rate is calculated from the autocorrelation of the heart based envelopes. The average heart rate and heart rate variability are then reported \cite{tang2021automated}.  
    \\
    \hline
    129-138
    & Percentage Bad Segmentation
    & Recordings are segmented using either the method proposed by Schmidt et al. or Springer et al \cite{springer2015logistic,schmidt2010segmentation}. Both methods used duration dependent hidden Markov models to separate the recording into four different states, namely, S1, systole, S2 and diastole. Segments assigned the same state are grouped together and outliers corresponding to poor segmentation are identified \cite{abdollahpur2017detection}. 
    \\
    \hline
    139-140
    & Segmentation Quality
    &  Recordings are segmented using either the method proposed by Schmidt et al. or Springer et al. \cite{springer2015logistic,schmidt2010segmentation}. For each S1 and S2 segments, 4 level MFCC decomposition is performed, which is re-sampled to length 14 and transformed into 70 (4 coefficients + log energy $\times$ 14) length cepstral vector. Total cepstral distances between all S1 segments (dS1) and all S2 segments are calculated (dS2). Segment quality is then represented as the combine total of ceptral distances (dS1 + dS2) \cite{beritelli2009heart}. 
    \\
    \hline
    141-146
    & Percentage Abnormal Segmentation
    & For Schmidt et al. and Springer et al. segmented signals, ZCR, RMSSD and SD1 of Poincaré plot are calculated for the systolic, diastolic segments separately and systolic and diastolic segments combined. The difference between systolic and diastolic segment values divided by combined segment is calculated. The percentage of segments above 0.8, 0.8 and 0.6 for RMSSD, SD1 and ZCR respectively is then calculated \cite{grzegorczyk2016pcg}. 
    \\
    \hline
    \end{tabular}
    \label{tab1:features}
\end{table*}

From our recent work, a total of 182 and 187 features for lung and heart sound quality classification are extracted \cite{grooby2020neonatal}. As strong high-quality heart sound can act as noise and reduce the quality of lung sound and vice versa, the heart and lung features were combined together and used for both heart and lung signal quality classification. 

The initial feature set was also expanded in 3 ways. Firstly, both 5\,s truncated and full autocorrelation signal of Hilbert envelope were used to calculate the autocorrelation-based features, as proposed by Springer et al. \cite{springer2016automated}. Secondly, Hilbert, homomorphic, Shannon, Short-Time Fourier Transform (STFT), power for 40-60\,Hz and 3\textsuperscript{rd}-level detailed wavelet coefficients with \textit{rbio3.9} wavelet envelopes were calculated as they commonly represent heart signals \cite{shannon2001mathematical, tang2021automated,liang1997heart,springer2015logistic}. Similarly, log-variance, variance fractal dimension, spectral energy and powers in 0-500\,Hz, 150-300\,Hz, 300-450\,Hz and 150-450\,Hz bands were calculated as they are commonly used to represent lung signals \cite{yap2001respiratory,huq2010automatic,huq2012acoustic,moussavi2000computerised}. As opposed to just using Hilbert envelope as in past work, all heart signal based envelopes were used to calculate hidden semi-Markov model (HSMM) quality features \cite{shi2019automatic}. 
Finally, the percentage acceptable windows feature was modified from our past work \cite{grooby2020neonatal}. Original feature used a sliding window of 2200ms with 25\% overlap and calculated the number of heart peaks within each window using a method proposed by Gieraltowski et al.\cite{gieraltowski2015rs}. The percentage of windows containing the normal range of 2-4 heart peaks was then calculated and used for to estimate signal quality \cite{gieraltowski2015rs,grzegorczyk2016pcg}. Percentage windows with number of peaks 4-7 and 2-8 were also considered, as these correspond to approximately the 5\textsuperscript{th} and 95\textsuperscript{th} percentile heart rate and full heart rate ranges of newborns, respectively \cite{Clinical22:online,fleming2011normal,grooby2020neonatal}. Additionally,  Springer et al., Schmidt et al., and Liang et al. methods were used to detect S1 and S2 heart peaks, as opposed to just heart peak detection using Gieraltowski et al. method previously. The percentage windows in acceptable ranges of 4-8 (original) 9-14 (5\textsuperscript{th}-95\textsuperscript{th} percentile) and 5-16 (full range) were then determined based on these detected S1 and S2 peaks \cite{springer2015logistic,schmidt2010segmentation,liang1997heart, grzegorczyk2016pcg,Clinical22:online,fleming2011normal,grooby2020neonatal}. The percentage acceptable windows features was also adapted for lung with 4~\,s sliding window with 25\% overlap and peak detection of inspiration and expiration peaks using methods developed in our past work \cite{grooby2020neonatal}. Percentage of windows with 1-4 and 1-5 corresponding to 5\textsuperscript{th} and 95\textsuperscript{th} and full range of respiratory rate, respectively where then calculated \cite{Clinical22:online,fleming2011normal,grooby2020neonatal}. 

146 additional features based on previous literature were also extracted, as summarized in Table~\ref{tab1:features}. Features 85-100 used lung-based envelopes and features 85-128 used the aforementioned heart-based envelopes.
In total 400 features were extracted for lung and heart sound quality classification. The source codes for these features are provided online in \cite{GitHubeg75:online}. 

\subsubsection{Feature Selection}
The training set was normalized to have zero means and unit variance, with these same scaling and shifting values used on the test set. 

For feature selection, the training set was class balanced with random up-sampling with replacement and maximum Relevance Minimum Redundancy (mRMR) algorithm with Mutual Information Difference (MID) method used \cite{peng2005feature}. mRMR maximizes relevance D (Equation \ref{eq:relevance}) and minimizes redundancy R (Equation \ref{eq:redundancy}) based on their difference (Equation \ref{eq:overall}), in a first order incremental search to rank most important features as calculated below:

\begin{equation}
\label{eq:relevance}
\max D(S,c), \quad D= \frac{1}{|S|}\sum_{x_i \in S}I(x_i;c)
\\
\end{equation}

\begin{equation}
\label{eq:redundancy}
\min R(S), \quad R=\frac{1}{|S|^2}\sum_{x_i,x_j \in S}I(x_i,x_j),
\\
\end{equation}

\begin{equation}
\label{eq:overall}
\max\phi(D,R), \quad\phi=D-R
\\
\end{equation}

Where S is the feature set, $x_i,x_j$ are individual features, I is mutual information, and c is the target class. 

The mean-square error was plotted against the number of features used based on mRMR algorithm in Figure \ref{fig:featureselection}. From this figure, heart classification performance plateaus from feature 5 onwards and lung classifier performance degrades after feature 20. To find region of best performance and minimize overfitting, the ranges of top 5-15 feature for heart and top 5-20 features for lung  were chosen for hyperparameter optimization.

\subsubsection{Classification}
The overall model is shown in Algorithm \ref{alg:model}, which takes in all recording features, patient assignment, signal quality annotations and hyper-parameters as input to train the classifier. 

As shown in Table \ref{fig:HeartLungAnnotationFiltered}, the distribution of signal qualities is not even. In particular, there are few recordings of high-quality, this is because recording in a neonatal intensive care environment is challenging with a large range of noises occurring. In order to resolve this, patient-wise class balancing was performed with the minority class being randomly upsampled with replacement.

Two groups of classifiers were implemented. The first group was standard regression methods that either had no parameters \{ordinary least squares regression, AdaBoost, gradient boosting, bagging and random forest\}, had regularisation strength optimised through 5-fold cross-validation \{ridge regression (alpha = 0.1, 0.5, 1.0, 5, 10.0, 50, 100, 500, 1000), LASSO, Elastic-Net ($l_1$ ratio=0.001, 0.005, 0.01, 0.05, 0.1, 0.5, 0.7, 0.9, 0.95, 0.99, 0.995, 0.999, 1.0), least angle regression, LASSO with least angle regression (max iterations=50) and orthogonal matching pursuit\} or had numerous parameters optimised using 5-fold cross-validation grid search parameter optimisation based on mean square error \{support vector machine, decision tree and k-nearest neighbours\} \cite{scikit-learn}. The second group was ordinal regression methods with either single parameter \{least absolute deviation (max iterations=5000)\} or had numerous parameters optimised using grid search (logistic model with all or immediate threshold, ridge and support vector machine) \cite{pedregosa2015feature}. 

With the standard regression method group, test set outputs were restricted to be in the range 1-5, whereas the ordinal regression is similar to multi-class classification, except that the order of the annotations is factored into the training. That is, if the correct output is 1, then a misclassification of 2 is better than 3. In fact, existing multi-class classifiers can be modified to be ordinal regression classifiers, which was done for support vector machine \cite{frank2001simple}.  

Patient-wise cross-validation was performed and all parameters tested in the grid search hyperparameter are listed below: 

\begin{algorithm}[tb]
\begin{flushleft}
\caption{Quality Assessment Model}
\label{alg:model}
\begin{algorithmic}[1]
\REQUIRE{$features,patientList,annotations,params$\;}
	\FOR{each $patient$ in $patientList$}
	\STATE $testSet$ $\gets$ $features(patient)$\;
	\STATE $testLabel$ $\gets$ $annotations(patient)$\;
	\STATE $trainSet$ $\gets$ $features(!patient)$\;
	\STATE $trainLabel$ $\gets$ $annotations(!patient)$\;
	\STATE $trainPatients$ $\gets$ $patientList(!patient)$\;
\ENDFOR

	\STATE $scaler$ $\gets$StandardScaler().fit($trainSet$)\;
	\STATE $trainSet$ $\gets$ $scaler.transform(trainSet)$\;
	\STATE $testSet$ $\gets$ $scaler.transform(testSet)$\;
	
	\STATE $trainSet\_balanced, trainLabel\_balanced$ $\gets$ RandomOverSampler($trainSet$, $trainLabel$) \;
	
	\STATE $topFeatures$ $\gets$ mRMR($trainSet\_balanced, trainLabel\_balanced$,`MID')\;
	
	\STATE $folds$ $\gets$ StratifiedCV($trainSet$, $trainLabel$,$trainPatients$,splits=5)\;
	
	\FOR{each $fold$ in $folds$}
	\STATE $fold$ $\gets$ RandomOverSampler($fold$)\;

	\STATE $R1$ $\gets$ GridSearch($Regressor(), params, mse$)\;
	\STATE $R1.fit(folds)$\;
	
	\STATE $R2$ $\gets$ RegressorCV($params$)\;
	\STATE $R2.fit(folds)$\;
\ENDFOR
\end{algorithmic}
\end{flushleft}
\end{algorithm}

\begin{itemize}
\item Support Vector Machine (SVM)
\begin{itemize}
\item Kernel= Radial basis function or linear kernel
\item Kernel Coefficient= 0.1, 0.01, 0.001, 0.0001, inverse of number of features, or inverse of number of features times variance
\item Regularization parameter C= 0.02, 0.04, 0.08, 0.16, 0.32, 0.64, 1.28, 2.56 or 5.12
\end{itemize}
\item Decision Tree (Tree)
\begin{itemize}
\item Measure quality of tree split=  Mean square error, Friedman mean square error, mean absolute error, Poisson deviance
\item Max depth of tree= Any, 1, 2, 3, 4, 5, 6, 7, 8, 9 or 10
\item Max features in each split= All features, square root of all features, or log2 of all features 
\end{itemize}
\item K-Nearest Neighbours (KNN)
\begin{itemize}
\item Number of neighbours= 1, 2, 3, 4, 5, 6, 7, 8, 9 or 10
\item Weight function= Uniform or inverse of distance
\item Algorithm to compute nearest neighbours= Brute-force search, BallTree or KDTree
\item Definition of distance=  Manhattan distance (l1) or euclidean distance (l2)
\end{itemize} 
\item Logistic Model with All or Immediate Threshold
\begin{itemize}
\item Alpha= 0, 0.1, 0.5, 1, 5, 10, 50, 100, 500, 1000
\item Max iterations= 10,000
\end{itemize}
\item Ordinal Ridge 
\begin{itemize}
\item Alpha= 0, 0.1, 0.5, 1, 5, 10, 50, 100, 500, 1000
\end{itemize}
\end{itemize}

\section{Heart Rate and Breathing Rate Error}
\label{sec:sync}
The purpose of this section is to analyse the relationship between heart signal quality and heart rate estimation error, and similarly for lung signal quality and breathing rate estimation error.

Using the 30 audio recordings with synchronous electrocardiogram, heart rate and breathing rate were automatically calculated every second with the inbuilt Dräger Infinity® M540 system algorithm \cite{Infinity28:online}. Electrocardiogram is considered the gold standard method for estimation of heart and breathing rate and is used as reference \cite{kevat2017systematic}. These recordings were not involved in training the regression model for quality assessment and were held out for testing. 

For the heart audio recordings, heart rate in beats per minute, was estimated every second with a sliding window of 3\,s. A sliding window of 3\,s was chosen as this is a sufficient length to obtain a minimum of 3 heart beats, necessary for accurate heart rate estimation. Two methods were used to estimate heart rate. Firstly, using method proposed by Schmidt et al. \cite{schmidt2010segmentation} where the autocorrelation of the Hilbert Envelope is calculated. The maximum peak is then detected in the autocorrelation signal between the bounds of 70-220 beats per minute. The range of 70-220 beats per minute is chosen as this is the typical heart rate range for newborns \cite{grooby2020neonatal,FastSlow12:online,Normalhe33:online}. 
The second method proposed by Springer et al. \cite{springer2015logistic}, uses the initial estimate of heart rate from the Schmidt et al. method as input into a duration-dependent hidden Markov model, to segment the heart beats to 4 states, namely S1, S2, systolic and diastolic.

For lung audio recordings, breathing rate in breaths per minute, was estimated every second with a sliding window of 6\,s. Similarly as before, a sliding window of 6\,s was chosen as this is a sufficient length to obtain a minimum of 3 breathing periods, which is necessary for accurate breathing rate estimation. For breathing rate estimation, power spectral envelope is calculated for the frequency range 300-450Hz and then peak detection is performed \cite{grooby2020neonatal}.

Using the regression quality assessment model proposed, this was trained using the heart and lung recordings that did not have synchronous electrocardiogram recordings. Heart and lung signal quality for the 30 synchronous recordings was then estimated using this trained model.

\begin{table*}[tb]
\caption{Heart rate and breathing rate errors}
    \centering
    \begin{tabular}{ |p{2cm}||
    p{0.8cm}|p{0.8cm}|p{0.8cm}|p{0.8cm}|p{0.8cm}||
    p{0.8cm}|p{0.8cm}|p{0.8cm}|p{0.8cm}|p{0.8cm}|}
    \hline
    &\multicolumn{5}{c||}{\textbf{Mean Absolute Error (bpm)$^{*}$}} 
    &\multicolumn{5}{c|}{\textbf{\% Acceptable$^{**}$}} 
    \\
    \hline
    \textbf{Signal Quality}
    & \textbf{1}
    & \textbf{2}
    & \textbf{3}
    & \textbf{4}
    & \textbf{5}
    & \textbf{1}
    & \textbf{2}
    & \textbf{3}
    & \textbf{4}
    & \textbf{5}
    \\
    \hline
    Heart Rate
    \newline Schmidt et al. 
    & 44.2
    & 18.2
    & 12.3
    & 5.2
    & 2.3
    
    & 26.8
    & 38.5
    & 70.3
    & 74.5
    & 92.6
    \\
    \hline
    Heart Rate
    \newline Springer et al. 
    & 49.3
    & 20.9
    & 13.8
    & 7.5
    & 4.9
    
    & 9.8
    & 15.4
    & 40.7
    & 49.0
    & 66.7
    
    \\
    \hline
    Breathing Rate
    & 30.6
    & 14.1
    & 12.0
    & 4.6
    & 2.8
    
    & 0
    & 9.1
    & 31.2
    & 66.7
    & 66.7
    \\
    \hline
    \multicolumn{11}{p{0.8\textwidth}}{$^{*}$bpm denotes beats or breathing periods per minute. Mean absolute error is calculated based on heart/breathing rate estimation method in comparison to gold standard synchronous electrocardiogram estimation.}\\
    \multicolumn{11}{p{0.8\textwidth}}{$^{**}$Acceptable \% refers to the proportion of the recordings in which heart/breathing rate error is less than 5 bpm}
    \\
    \end{tabular}
    \label{tab1:hrbr_error}
\end{table*}

\section{Real Time Processing}
\label{sec:real}
\begin{figure*}
     \centering
    \begin{subfigure}[b]{0.48\textwidth}
         \centering
         \includegraphics[scale=0.55,trim={0.5cm 0.3cm 0.5cm 0.5cm}]{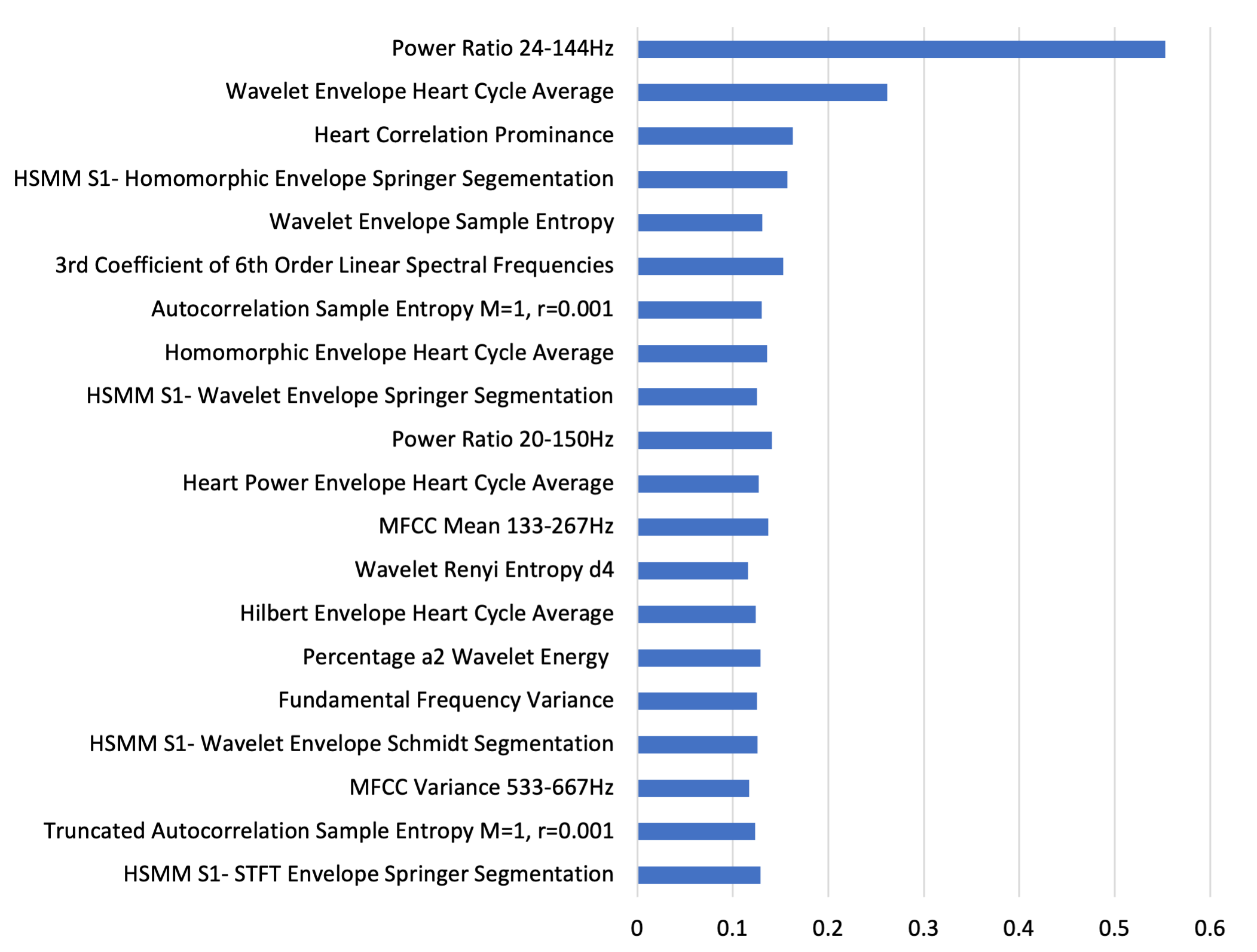}
         \caption{Heart}
     \end{subfigure}
     \begin{subfigure}[b]{0.48\textwidth}
         \centering
         \includegraphics[scale=0.55,trim={0.5cm 0.3cm 0.5cm 0.5cm}]{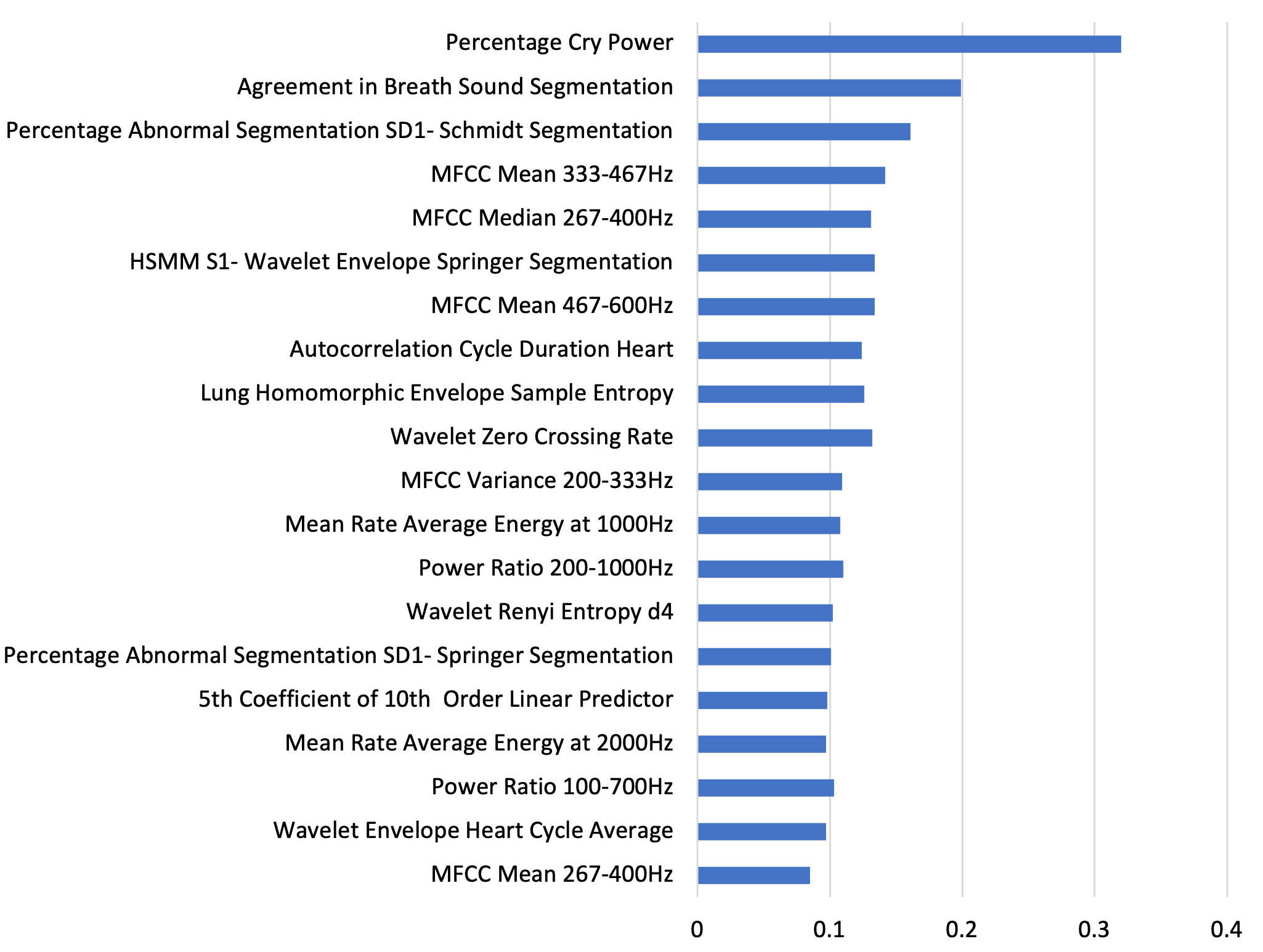}
         \caption{Lung}
     \end{subfigure}     
        \caption{The top 20 features based on mRMR algorithm with MID method \cite{peng2005feature}}
        \label{fig:featureimportance}
\end{figure*}

The top 20 features based on mRMR algorithm are shown in Figure \ref{fig:featureimportance}. Median time for feature extraction was calculated using MATLAB 2021a with MacBook Pro CPU 2.3\,GHz 8-Core Intel~i9. For extracting all 20 features, 1.46\,s and 2.16\,s is required for heart and lung, respectively. The best performing classifiers used 15 and 19 features for heart and lung, which corresponded to 1.12 seconds and 2.16\,s, respectively. 

Time consuming features to calculate were:
\begin{itemize}
\item Sample entropy of autocorrelation signal which takes 850ms and 210ms for full and 5\,s truncated autocorrelation signal. 
\item Heart segmentation based features as Schmidt et al. and Springer et al. segmentation take 120\,ms and 80\,ms, respectively \cite{springer2015logistic,schmidt2010segmentation}
\item Heart and lung-based singular value decomposition, which take between 50-250ms per feature
\item STFT envelope-based features, as STFT envelope takes 120\,ms to calculate
\item Mean rate average energy at 1000\,Hz and 2000\,Hz that take 640\,ms and 1.28\,s to calculate
\end{itemize}

\begin{figure*}
     \centering
    \begin{subfigure}[b]{0.48\textwidth}
         \centering
         \includegraphics[scale=0.55,trim={0.5cm 0.3cm 0.1cm 0.5cm}]{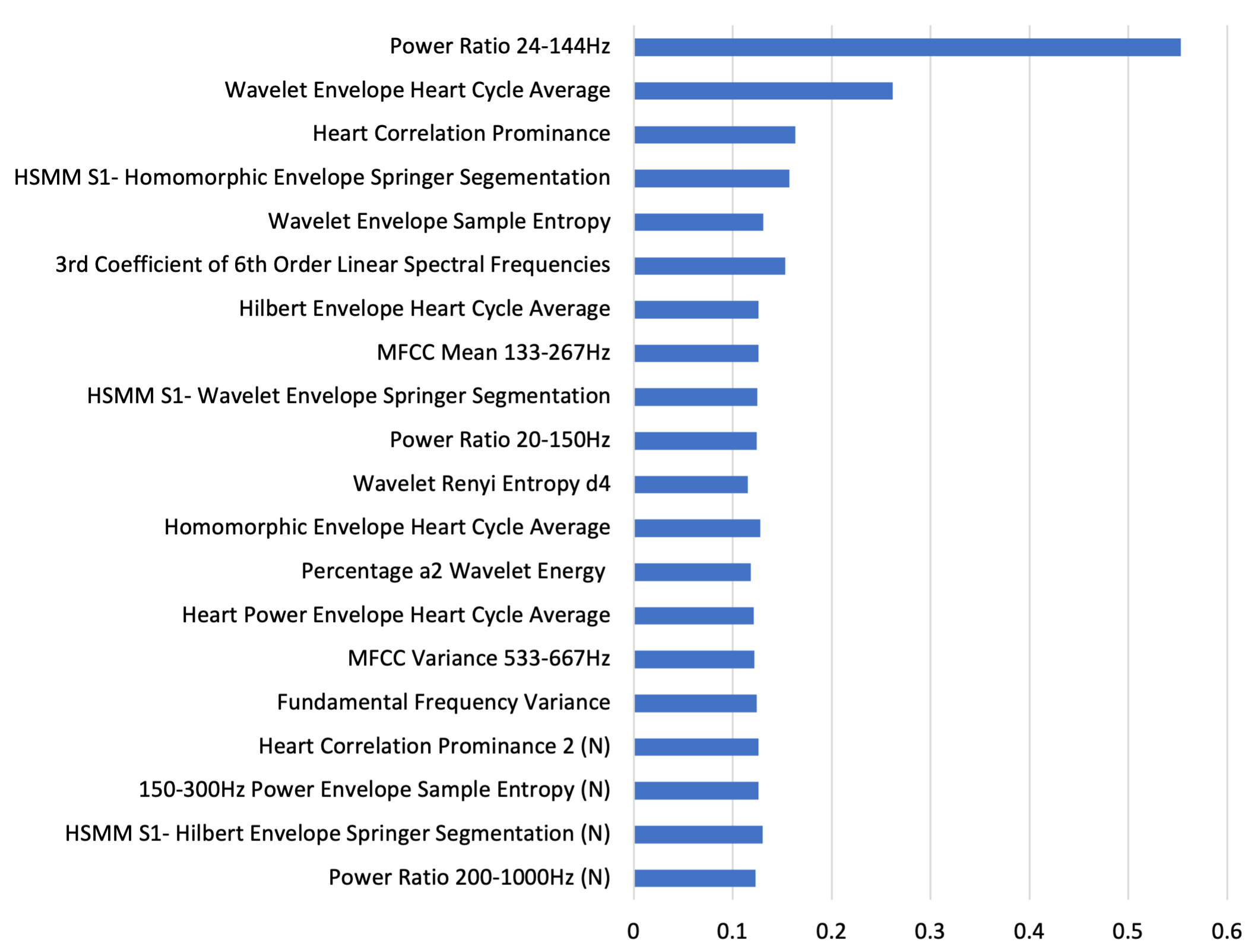}
         \caption{Heart}
     \end{subfigure}
     \begin{subfigure}[b]{0.48\textwidth}
         \centering
         \includegraphics[scale=0.55,trim={2.5cm 0.3cm 0.5cm 0.5cm}]{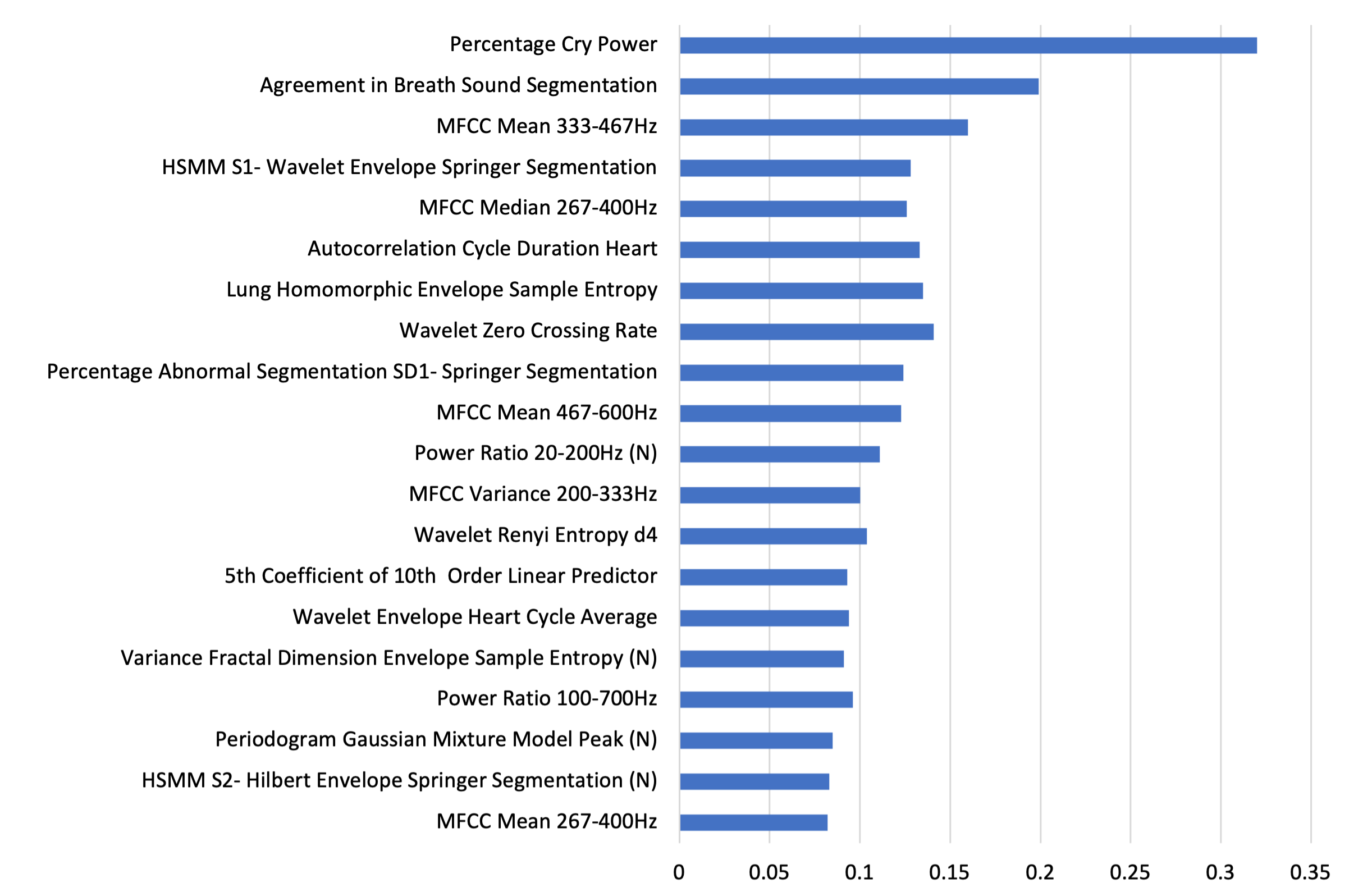}
         \caption{Lung}
     \end{subfigure}     
        \caption{The top 20 fast features calculated based on mRMR algorithm with MID method \cite{peng2005feature}. The slow features have been removed as detailed in Section~\ref{sec:real}. New features not seen in Figure \ref{fig:featureimportance} are labelled (N).}
        \label{fig:featureimportancefast}
\end{figure*}

All features above were removed, except for Springer et al. segmentation, as many of the top features in both heart and lung utilised segmentation based features. Figure \ref{fig:featureimportancefast} shows the new top 20 features, which take 160\,ms and 200\,ms to extract for heart and lung, respectively. The best performing classifiers used 13 and 14 features for heart and lung, which corresponded to 130ms and 120ms, respectively. 

\begin{table*}[tb]
    \begin{center}
\caption{Summary of classifier results. Heart and Lung Classifiers are trained with all features, whereas Heart Fast and Lung Fast Classifiers have slow features removed (cf. Section \ref{sec:real}).}
    \begin{tabular}{ |p{1.2cm}||p{1.2cm}|p{1.2cm}|p{1.2cm}|p{1.2cm}||p{1.2cm}|p{1.2cm}| }
    \hline
    \textbf{Classifier}
    & \textbf{Test \newline MSE$^*$}
    & \textbf{Train \newline MSE}
    & \textbf{Test \newline Acc (\%)}
    & \textbf{Train \newline Acc (\%)}
    & \textbf{Test \newline BAcc (\%)}
    & \textbf{Train \newline BAcc (\%)}\\
    \hline
    Heart
    & 0.487
    & 0.247
    & 54.7
    & 74.6
    & 56.8
    & 75.6
    \\
    \hline
    Heart Fast
    & 0.459
    & 0.272
    & 55.0
    & 72.5
    & 56.7
    & 74.0
    \\
    \hline
    Lung
    & 0.612
    & 0.207
    & 54.4
    & 81.5
    & 51.2
    & 81.7
    \\
    \hline
    Lung Fast
    & 0.673
    & 0.219
    & 47.9
    & 81.1
    & 46.3
    & 81.1
    \\
    \hline
    \multicolumn{7}{l}{$^*$MSE: Mean Squared Error, Acc: Accuracy, BAcc: Balanced Accuracy.}\\    
    \end{tabular}
    \label{tab1:regressionresults}
\end{center}    
\end{table*} 
 
\begin{figure*}
     \centering
     \begin{subfigure}[b]{0.48\textwidth}
         \centering
         \includegraphics[scale=0.34,trim={0.5cm 0.5cm 0.5cm 0.5cm}]{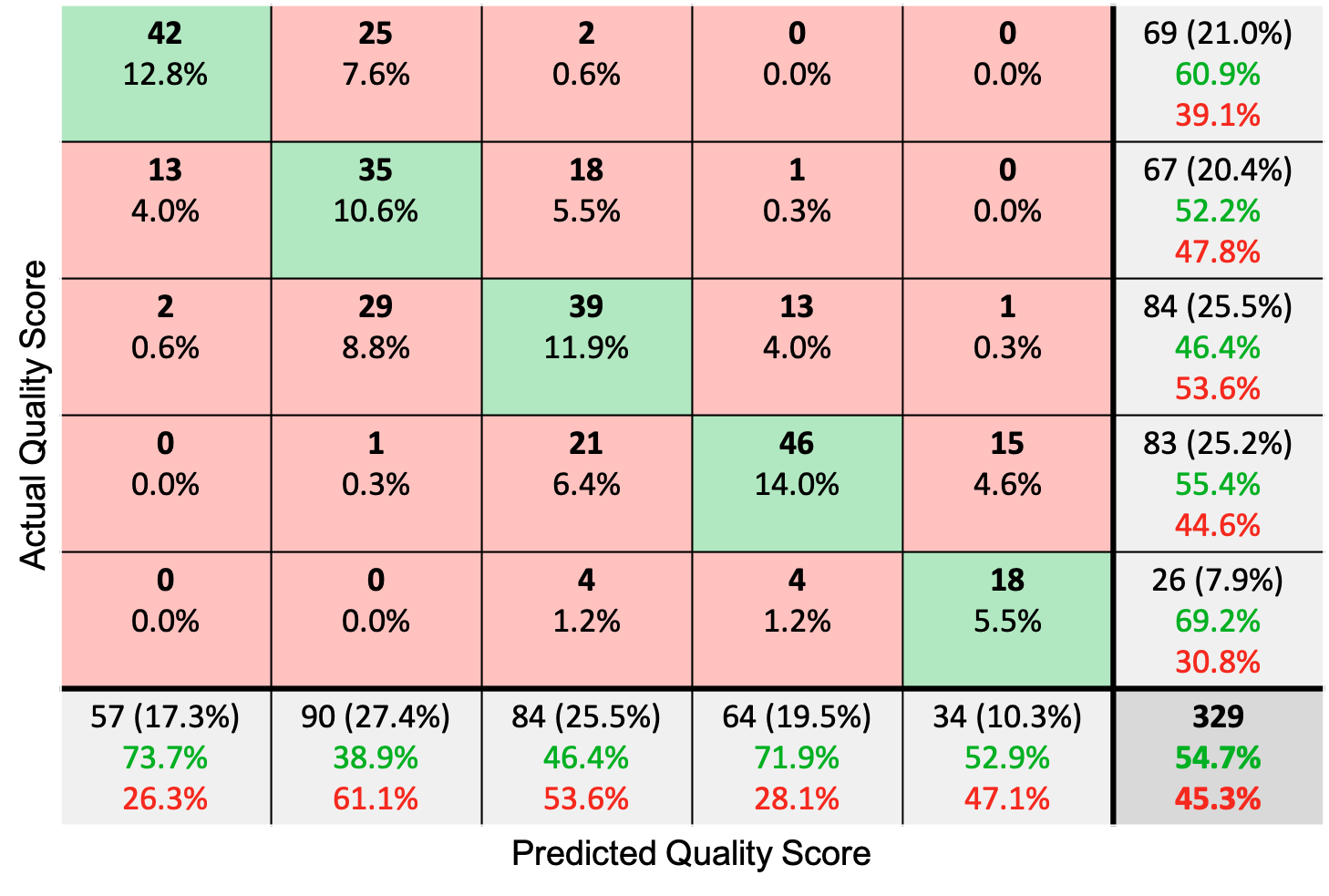}
         \caption{Heart}
     \end{subfigure}
     \begin{subfigure}[b]{0.48\textwidth}
         \centering
         \includegraphics[scale=0.34,trim={0.5cm 0.5cm 0.5cm 0.5cm}]{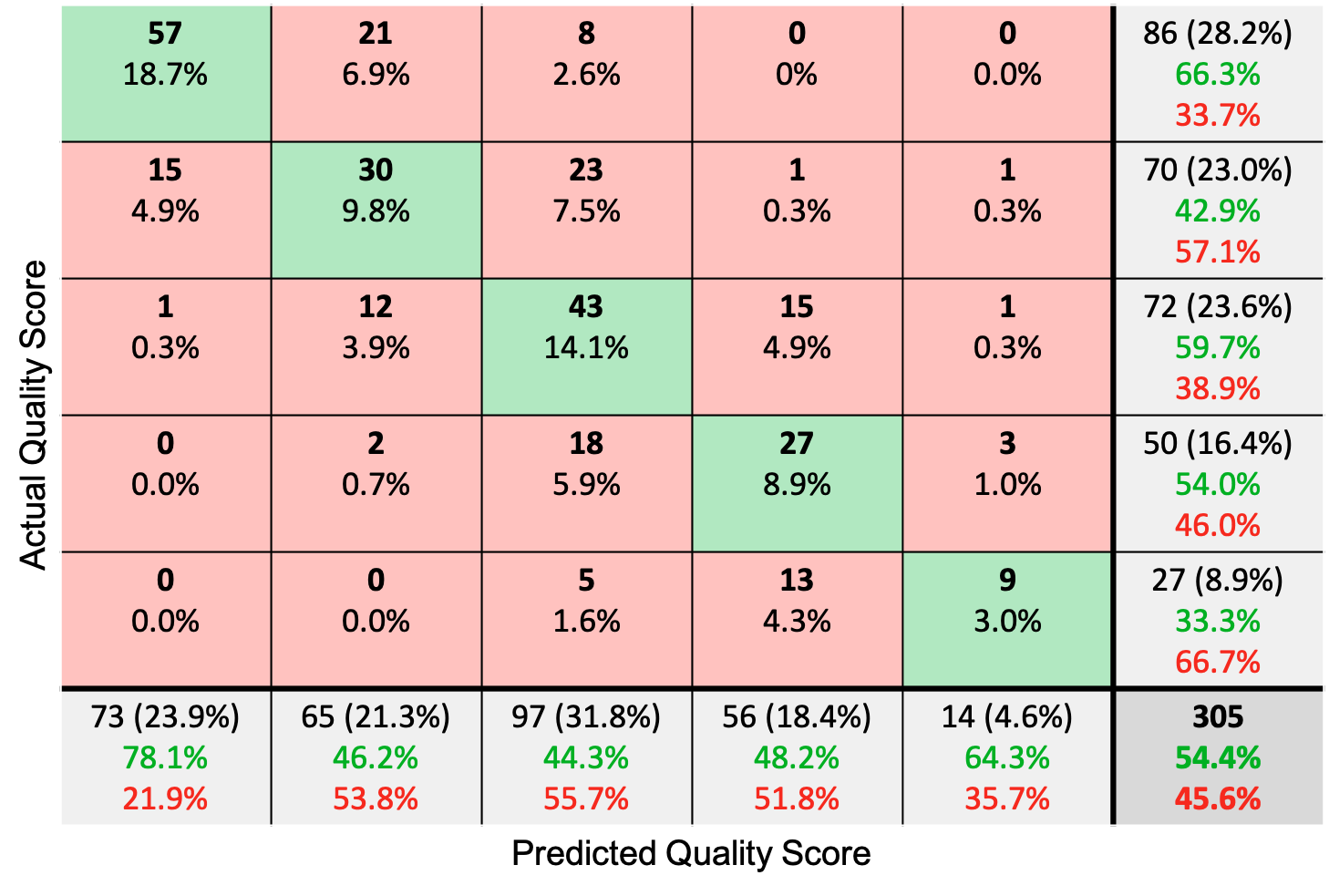}
         \caption{Lung}
     \end{subfigure}
        \caption{Confusion matrix of signal quality estimation}
        \label{fig:confusionresults}
\end{figure*}

\begin{figure}[tb]
\centering
\includegraphics[scale=0.34,trim={0.5cm 0.5cm 0.5cm 0.5cm}]{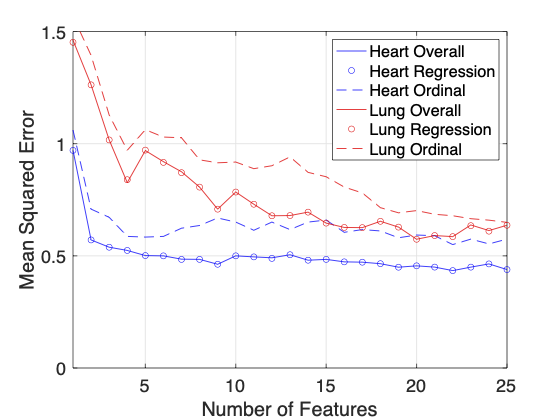}
\caption{Top features utilised vs MSE. The top features are based on mRMR algorithm with MID method \cite{peng2005feature}. Classifiers are grouped as either regression based (circle) or ordinal regression based (dashed line) with results shown for heart (blue) and lung (red). Solid lines show the best performing classifier result for each feature value.}
\label{fig:featureselection}
\end{figure}

\section{Results}
\label{sec:results}

Figures \ref{fig:featureimportance} and \ref{fig:featureimportancefast} show the top 20 features with all 400 features and slow features removed, respectively. Corresponding classifier results based on these top features are shown in Table \ref{tab1:regressionresults}. Heart with and without slow features perform comparably with balanced accuracy across the 5 classes being 56.8\% and 56.7\%, respectively. On the other hand, the removal of slow lung features results in a noticeable decrease in performance in all categories (mean squared error, accuracy and balanced accuracy). 

\begin{figure*}
     \centering
     \begin{subfigure}[b]{0.31\textwidth}
         \centering
         \includegraphics[scale=0.31,trim={0.5cm 0.5cm 0.5cm 0.5cm}]{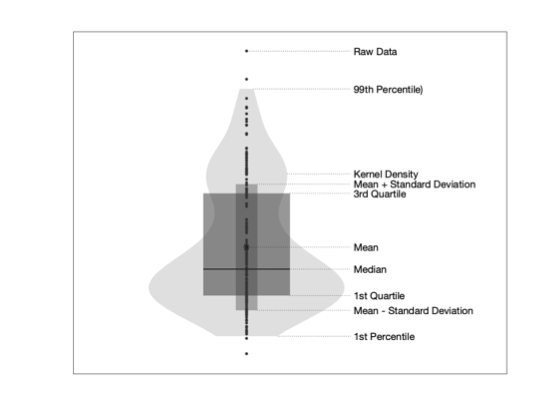}
         \caption{Legend}
     \end{subfigure}
     \begin{subfigure}[b]{0.31\textwidth}
         \centering
         \includegraphics[scale=0.31,trim={0.5cm 0.5cm 0.5cm 0.5cm}]{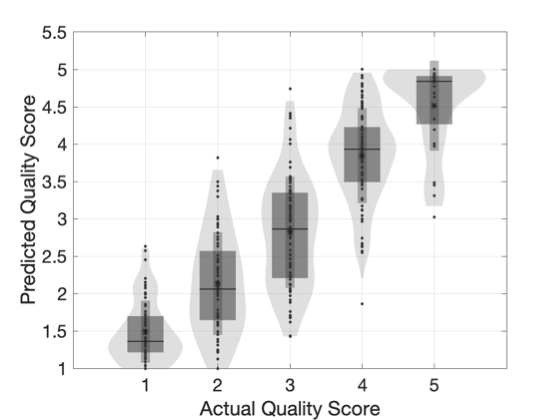}
         \caption{Heart}
     \end{subfigure}
     \begin{subfigure}[b]{0.31\textwidth}
         \centering
         \includegraphics[scale=0.31,trim={0.5cm 0.5cm 0.5cm 0.5cm}]{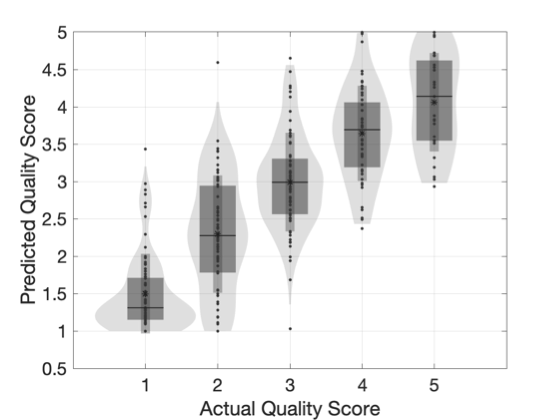}
         \caption{Lung}
     \end{subfigure}     
        \caption{Violin plot of signal quality estimation}
        \label{fig:violinresults}
\end{figure*}

Patient-wise cross-validation results using top 5-15 heart and 5-20 lung features are shown in Figures \ref{fig:confusionresults} and \ref{fig:violinresults}. Distinct separation of classes can be observed in Violin plots, however, there is a large overlap between classes due to 25-75th percentile generally varying +/-0.5 from the median. This overlap between is further supported in the confusion matrix results, with estimated signal quality concentrated +/-1 in class and the observed accuracy of 54.7\% and 54.4\% for heart and lung, respectively. Top features utilised vs mean square error are shown in Figure \ref{fig:featureimportance}. Best performing standard regression models outperforms ordinal regression models at all number of features based on both mean squared error and accuracy. 

The regression quality assessment model used for heart and breathing rate estimation had a mean squared error of 0.505 and 0.742 and accuracy of 56.3\% and 43.8\% for heart and lung signal quality estimation.

Table \ref{tab1:hrbr_error} shows the mean absolute error and percentage of recordings with error less than 5bpm for signal qualities 1 to 5. As can be seen in all cases, improvement in signal quality leads to reduction in heart and breathing rate error and increase in the percentage of recordings with less than 5bpm error. For clinical use, mean absolute error of less than 5bpm is typically required \cite{springer2014robust,nizami2018measuring}. Based on this requirement, only high-quality recordings with signal quality 5 for heart recordings and signal qualities 4 and 5 for lung recordings meet this requirement. Whereas, low-quality recordings are not appropriate for accurate vital sign estimation.

\section{Discussion}
\label{sec:discussion}

In terms of modelling signal quality data, three options were available: multi-class classification, ordinal regression, and regression. As ordinal regression is a multi-class classification model that treats the classes as an ordered set that is consistent with signal quality labels, it was chosen over multi-class classification. However, knowing whether ordinal regression or standard regression is more appropriate is a more difficult task. Rationale for standard regression is signal quality makes sense as a continuous scale from 1-5 as noise volume and contamination can vary continuously. Furthermore, standard regression aids in addressing annotator disagreement shown in Figures \ref{fig:HeartLungAnnotationOriginal} and \ref{fig:HeartLungAnnotationOriginal}. Consider 2 recordings, both with median signal quality score of 5, but for one all 5 annotators scored the recording 5, whereas only 3 did for the other. This annotator disagreement suggest the former recording is of higher quality even though both are represented by the same score. While ordinal regression could only model in discrete classes, the benefit of standard regression of these two recordings can be scored differently more appropriately representing the actual signal quality. 

However, two issues arise from using standard regression. Firstly, signal quality estimation can go outside the range 1-5. This issue partially addresses with signal quality being restricted to 1-5 after classification, however, this does not change the inherent method used for training the classifier itself. Secondly, whilst signal quality makes sense to be represented as continuous value, this does not mean the discrete classes used for annotating are equally spaced. For instance, lung signal quality classes 4 and 5 sound closer to each other than signal quality classes 1 and 2 as demonstrated in Figure \ref{fig:violinresults}. This makes sense, as no or next to no lung sound (class 1) vs hearing partially lung sound (class 2) is an easier task than to differentiate easy to hear lung sound that both have minimal noise typically in the form of heart sound (class 4 and 5). As shown in Figure \ref{fig:violinresults}, this issue is addressed, but it means that signal quality between 1-5 is not completely evenly distributed. 

Based on Figure \ref{fig:featureselection} standard regression clearly outperformed ordinal regression based on mean squared error. This suggests that standard regression more appropriately represented signal quality, which fits the earlier discussion. Other contributing factors to superior performance is that continuous valued estimation is easier to minimize mean squared error, and a lager set of regression models in comparison to ordinal regression were available. With regards to lager set of regression models, in both python and MATLAB regression libraries, are more established, optimised and available, whereas fewer ordinal regression models are available. There is a simple method of converting several multi-class classification classifiers into a ordinal classifier, however, this is not ideal as training multiple classifiers independently is an inefficient process, and the potential for specialized algorithms that can train with a single classifier may produce superior results \cite{frank2001simple}.

As shown in Table \ref{tab1:hrbr_error}, high-quality (signal quality of 4 or 5) can enable accurate vital sign estimation of heart and breathing rate for clinical usage. Whereas, low-quality recordings can provide inaccurate vital sign estimation, hindering clinical diagnosis. It can also been seen that mean absolute error increases using the Springer et al. heart rate method in comparison to Schmidt et al. method. As Schmidt et al. method is used as an initial vital sign estimation for the Springer et al. method's heart segmentation, the increased error suggests that poor heart rate estimation amplifies the error in the more detailed analysis of heart segmentation. Overall, improvement in signal quality can enable more accurate vital sign estimation which is necessary for clinical use and more detailed analysis.

For real-time processing, slow features namely sample entropy to autocorrelation signal, mean rate average energy and features based on Schmidt et al. heart segmentation, STFT envelope, singular value decomposition features were removed. The removal of features meant feature extraction times were markedly reduced from 1.12\,s to 130\,ms and 2.16\,s to 120\,ms for heart and lung respectively. Real-time processing is less than 400\,ms processing time, which is satisfied with both classifiers; however, these processing times were achieved with a MacBook Pro \cite{valderrama2018improving}. Similar results would be expected if a desktop computer in a hospital setting or phone connected to cloud computing, whereas using phone onboard processing would be expected to be slower. Future research in investigating processing time on phones would be required to determine appropriateness. Reducing the processing times of 130\,ms and 120\,ms even further are possible. Most promising methods for reducing processing time is converting MATLAB code into optimised C code in MEX function and the other is vectorizing for loops.

For heart sound quality classification, the removal of slow features resulted in only minor changes in results (Table~\ref{tab1:regressionresults}). As only maximum of 15 features were used, only autocorrelation sample entropy feature was removed, which had comparable feature selection score to other features in the top 20, meaning the removal of that feature was minor. Furthermore, the removed features important for heart classification had analogous faster features, namely, downsampled sample entropy instead of autocorrelation sample entropy, Springer et al. method instead of Schmidt et al. heart segmentation and numerous other envelope representations instead of STFT envelope. Finally, as heart results improvement plateaued after 5 features (Figure \ref{fig:featureselection} the removal of features ranked 7, 17, 19, 20 would be expected to be minor. 

It is noted that in Table \ref{tab1:regressionresults} heart results with removed features perform slightly better with regards to accuracy and mean squared error , which may appear counter-initiative. Firstly, these differences are minor and secondly, as the classifiers were trained with balanced classes, comparison based on balanced classes would be more appropriate. When comparing heart results with removed features with regards to balanced accuracy, it performed slightly worse as expected. 

For breath sound quality classification, the removal of slow features produced a marked decrease in performance as shown in Table \ref{tab1:regressionresults}. This can be explained by the combination of higher ranked features 2, 12, 17 being removed and up to maximum of 20 features being used for the classifier as opposed to the heart classifier where lower ranked features were removed and only top 15 were features used. Additionally, their are not any features that closely resemble the removed mean rate average energy features that were removed \cite{kala2020}. Future works in optimising mean rate average energy for real-time processing can potentially address this decrease in performance. 

Heart and lung quality classifier performance achieved accuracy of 54.7\% and 54.4\% and mean squared error of 0.487 and 0.612, respectively. One reason for the relatively low accuracy can be attributed to the annotator disagreement with median annotated quality differing from typically by +/-1 by some annotators. This annotator disagreement can be seen in Figure \ref{fig:HeartLungAnnotationOriginal}, where accuracy was 64.3\% and 62.3\% for heart and lung, respectively. Whilst removal of poor agreement recordings was done which improved annotator accuracy to 71.2\% and 71.9\% for heart and lung, respectively, this still suggests there is difficulty in accurately defining signal quality. In particular, annotator disagreement is high for the middle classes 2-4, which is also observed in the classifier results in Figure \ref{fig:violinresults}. This suggests that while annotators can generally agree on what is clearly low and high-quality, middle values are a lot harder to determine. 

Potential solutions to address annotator disagreement to improve classification accuracy is the generation of artificial dataset of heart and lung sounds with varying levels of noise. More precisely, clean heart, lung, and variety of sources of noise such as stethoscope movement, alarms, crying and background talking can be fused together in different combinations. The signal quality label for these artificial recordings would then be the signal to noise ratio. Key benefit of the artificial dataset is clear definition of signal quality and the generation of a balanced large number of examples of varying signal quality for training of classifiers. However, a key question of this method is how closely these artificial recordings resemble real low and high-quality heart and lung sounds, and is their a strong correlation between signal to noise ratio and perceived signal quality by clinicians. Additional issue for the construction of artificial dataset is obtaining enough clean heart and lung sounds, in particular lung sounds as a majority are contaminated with heart noise. Regardless, there has been large amount of research in artificial heart/lung recording datasets mixed either instantaneously or via convolution \cite{kala2020,shah2014blind,lin2013blind,vincent2006performance,rudnitskii2014using,ghaderi2011localizing}.

Another contributing factor to relatively low accuracy for estimating signal quality is the small imbalanced training set. In particular, classification accuracy was only 33.3\% for class 5 in lung classifier, which contained only 8.9\% of the training data. As discussed previously, given that signal quality class 4 and 5 in lung appear to be difficult to differentiate (Figure \ref{fig:violinresults}) and are underrepresented in the dataset, a larger number of recordings for training would be of benefit. Larger number of recordings would also enable a larger feature set to be utilized before over-fitting becomes a major concern. Imbalanced classes was partially addressed with data up-sampling with replacement, however, this only replicates existing recordings which can lead to over-fitting problem.  Synthetic Minority Oversampling Technique (SMOTE) may address this by generating new samples from the underrepresented class based by interpolation from the existing recordings \cite{chawla2002smote}. As this is typically achieved with k=5 for k-nearest neighbors (KNN) in the underrepresented class in the features space, this is not viable with the current dataset. As 5-fold cross-validation is performed, their are some folds with fewer than 5 recordings. Furthermore, with only a small number of recordings, very few new interpolated recordings could be generated. 

Future collection of high-quality heart and lung sounds may address class imbalance and improve classifier performance. However, obtaining such recordings is difficult in noisy neonatal intensive care unit environment. One option to address this issue is the usage of more advanced denoising and sound separation techniques as opposed to standard frequency filtering. These methods can enable high-quality heart and lung sounds to be generated from noisy chest sound recordings. Non-negative matrix co-factorisation is one such method developed in previous work \cite{grooby2021new}. 

Similar to past work, heart classifier performance was superior to lung classifier performance as shown in Table \ref{tab1:regressionresults} \cite{grooby2020neonatal}. Annotator agreement both before and after removal of recordings was consistent for heart and lung, suggesting classification of signal quality is of similar difficulty. As majority of the features are either heart-based or been adapted from heart features, this resulted in more suitable features for classification of heart sound quality. Currently, there are fewer works in lung sound quality estimation, but as shown in Figures \ref{fig:featureimportance} and \ref{fig:featureimportancefast}, tailored features of agreement in breath sound segmentation, breath sound envelope and mean rate average energy are important for lung signal quality estimation. Therefore, future work in creating further lung-based features may improve results. 

With top features for signal quality classification, features with frequency ranges of 20-267Hz and 200-467Hz were observed for heart and lung sounds quality classifiers, respectively. This makes sense as their frequency ranges correspond closely with frequency ranges for heart and lung sounds \cite{grooby2020neonatal}. Additionally, many heart segmentation-based features such as HSMM quality and percentage abnormal segmentation features were important for lung sound classification. As heart sounds are normally present in all lung recordings and act as noise reducing signal quality, these heart segmentation-based features aid in the determining the amount of heart noise contamination.

\section{Conclusion}
\label{sec:conclusion}
Stethoscope-recorded chest sounds provide affluent information about neonatal health status, in particular for cardio-respiratory health assessment. In combination with telehealth, digital stethoscopes can increase the availability of quality healthcare for early diagnosis and prognosis of newborns. However, as shown in this paper, acquisition of high-quality recordings is necessary to obtain accurate vital signs for clinical use. In order to achieve this, accurate signal quality assessment is required for both heart and lung sounds recorded from the digital stethoscope. Signal quality assessment enables feedback to non-expert users on the quality of recordings and to aid the clinical decision support system for automated analysis of those recordings. This paper presented a newborn-focused automatic heart and lung sound quality assessment on a five-level quality scale using a variety of regression methods. Overall, for the best-performing classifiers, heart and lung quality were estimated with a mean squared error of 0.487 and 0.612, taking 1.12\,s and 2.16\,s to compute per recording, respectively. For real-time application, heart and lung quality were estimated in under 130ms with mean square error of 0.459 and 0.673 , respectively.  

\section*{Acknowledgment}

E. Grooby thanks and acknowledges the support from the Monash Newborn team at Monash Children's Hospital, Australia for data collection and Jinyuan He for the analysis of the newborn data in previous works. 

\bibliographystyle{IEEEtran}
\bibliography{IEEEabrv,references}

\begin{IEEEbiography}[{\includegraphics[height=1.25in,clip,keepaspectratio]{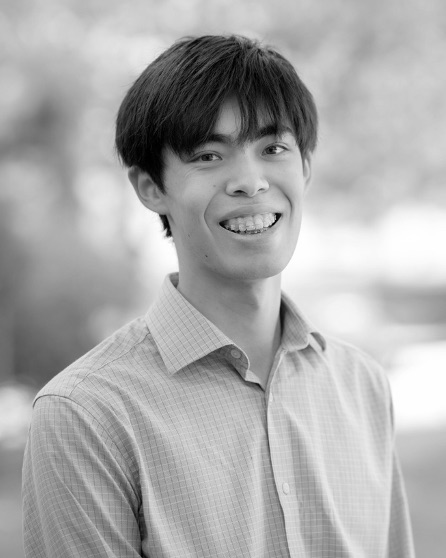}}]{E. Grooby} (M 2020) received degrees in Bachelor of Biomedicine in 2017 and Master of Engineering (Biomedical) in 2019 at The University of Melbourne, Melbourne, Victoria, Australia. He is currently pursuing a joint Ph.D. degree in electrical and computer systems engineering at Monash University, Melbourne, Victoria, Australia and The University of British Columbia, Vancouver, British Columbia, Canada. From 2017-2019, he was a Research Student at Walter and Eliza Hall Institute of Institute of Medical Research, Peter MacCallum Cancer Centre and Defence Science and Technology Group. In 2019-2020, he was a Research Engineer at Cochlear. His research interests include biomedical signal processing and medical device development. 
\end{IEEEbiography}

\begin{IEEEbiography}[{\includegraphics[height=1.25in,clip,keepaspectratio]{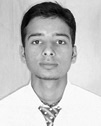}}]{C. Sitaula} 
 is a research fellow under the department of electrical and computer systems engineering, Monash University, Victoria, Australia. He received a Ph.D. degree from Deakin University, Victoria, Australia in 2021. He worked in industry and academia in Nepal for a number of years before joining Deakin University for his PhD degree. He has published research articles in top-tier conferences and journals in the field of deep learning and machine learning. His research interests include computer vision, signal processing, and machine learning.
\end{IEEEbiography}

\begin{IEEEbiography}[{\includegraphics[height=1.25in,clip,keepaspectratio]{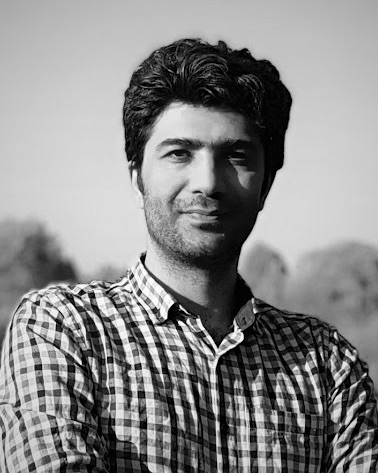}}]{D. Fattahi} is a PhD candidate in Biomedical Engineering (Bioelectric) at Shiraz University, Shiraz, Iran. He spent a research internship (Mar-Sep 2019) at Biomedical Signal Processing Lab in the department of electrical and computer systems engineering, Monash University, Australia. His recent fields of study include biomedical signal processing (ECG, EEG, PCG, lung sound), parameter estimation, time-frequency analysis and blind source separation.
\end{IEEEbiography}

\begin{IEEEbiography}[{\includegraphics[height=1.25in,clip,keepaspectratio]{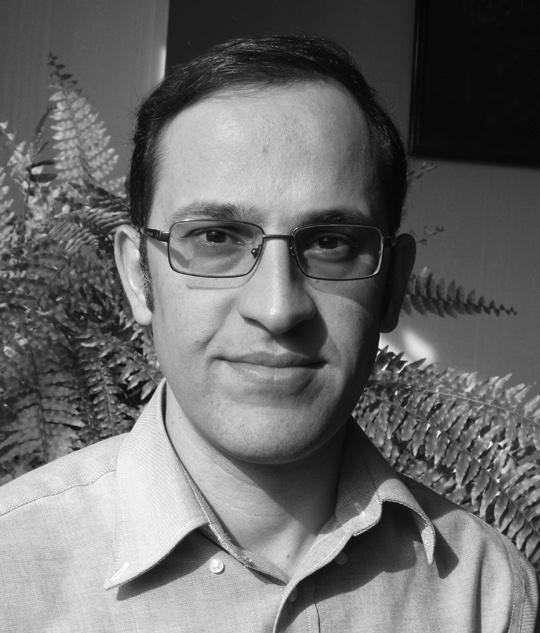}}]{R. Sameni} (S'2001-M'2009-SM'2015)  received a bachelor’s degree in Electronics Engineering from Shiraz University, Iran (2000), a master’s degree in Biomedical Engineering from Sharif University of Technology, Iran (2003), and a double Ph.D. degree in Signal Processing and Biomedical Engineering from Institut National Polytechnique de Grenoble (INPG), France, and Sharif University of Technology (2008). He was a tenured Associate Professor of the School of Electrical and Computer Engineering, Shiraz University (2008-2018), an invited senior researcher at GIPSA-lab, Grenoble, France (2018-2020), and is currently an Associate Professor of Biomedical Engineering at Emory University, GA, US (since 2020). Dr Sameni's research interests include statistical signal processing with special interest in mathematical modeling and analysis of biomedical systems and signals. 
\end{IEEEbiography}

\begin{IEEEbiography}[{\includegraphics[height=1.25in,clip,keepaspectratio]{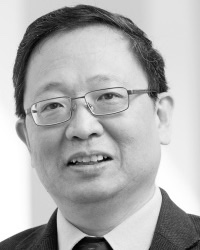}}]{K. Tan} is a Consultant Neonatologist at Monash Children’s Hospital and Adjunct Associate Professor at Monash University. He has research interests in clinical registry, clinical trials, big data analysis and clinical practise improvement.
\end{IEEEbiography}

\begin{IEEEbiography}[{\includegraphics[height=1.25in,clip,keepaspectratio]{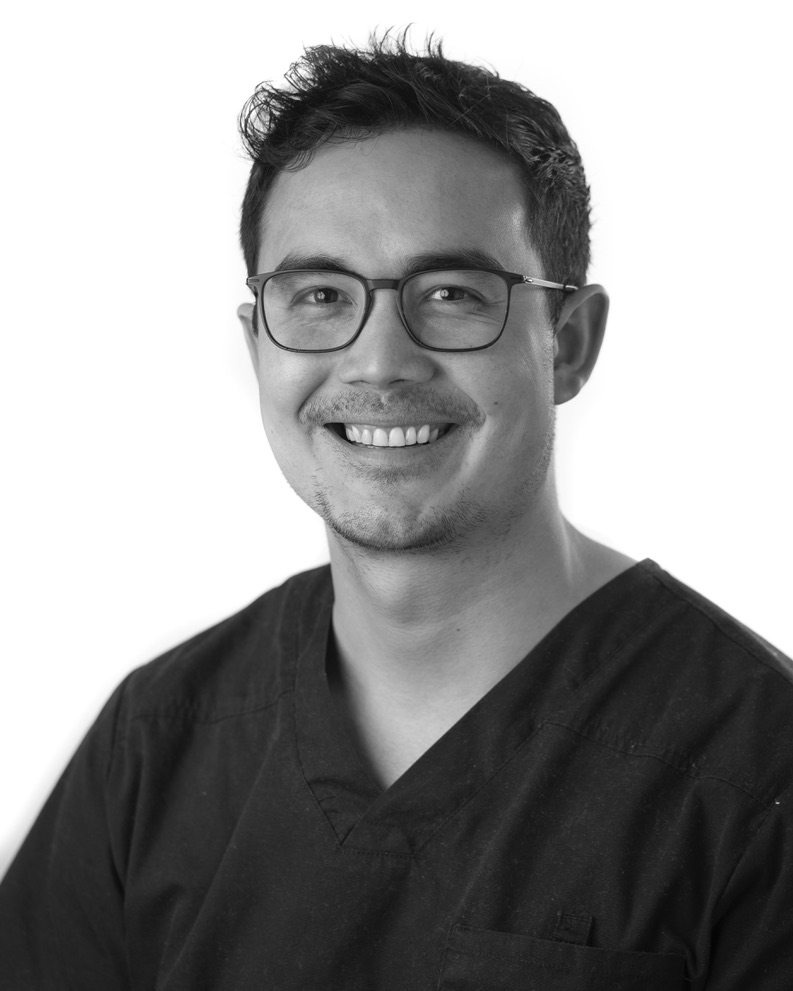}}]{L. Zhou} is a neonatologist at Monash Children’s Hospital, Melbourne, Australia, completing his neonatal fellowship training in 2020, and MBBS in 2012.
He is undertaking a PhD investigating umbilical cord blood-derived cell therapies for preterm brain injury at Monash University.
\end{IEEEbiography}

\begin{IEEEbiography}[{\includegraphics[height=1.25in,clip,keepaspectratio]{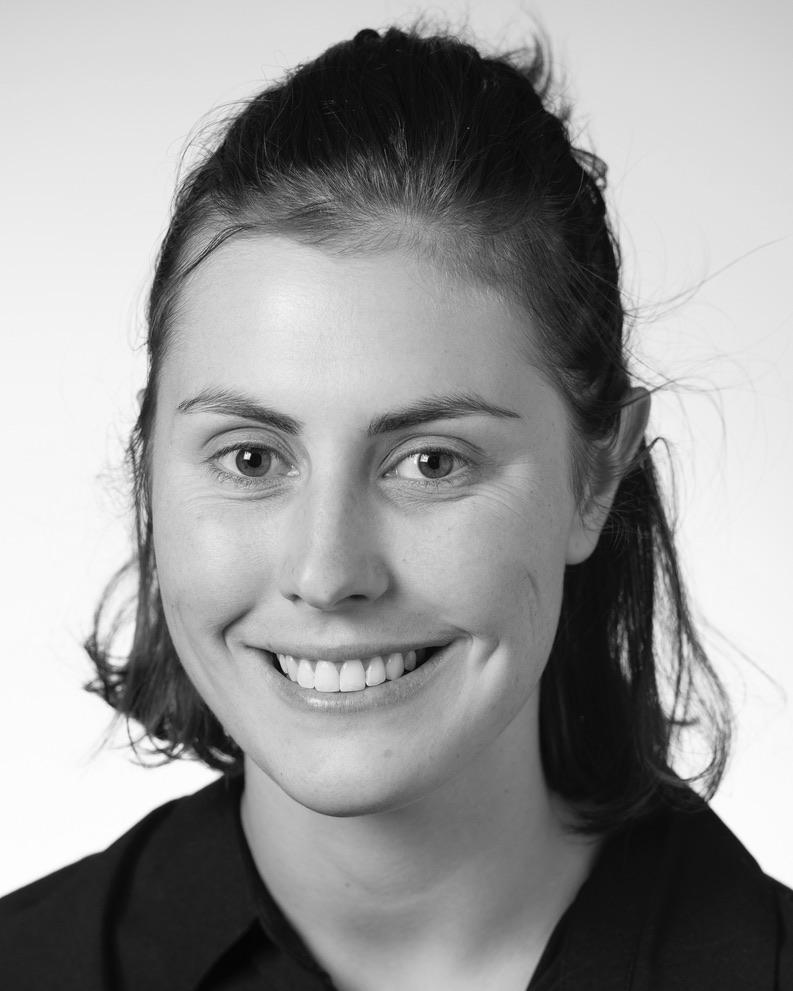}}]{A. King} received degrees in Bachelor of Medical Science (Hons) in 2019 and Bachelor of Medical Science and Doctor of Medicine in 2020 through Monash University, Melbourne, Australia. She is working as a junior doctor at St Vincent’s Hospital, Melbourne, Australia and holds an affiliate position within the Department of Paediatrics, Monash University. 
\end{IEEEbiography}

\begin{IEEEbiography}[{\includegraphics[height=1.25in,clip,keepaspectratio]{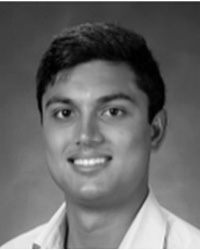}}]{A. Ramanathan} is a Resident Medical Officer at Perth Children’s Hospital and has received degrees in Bachelor of Medical Science (Hons) and MBBS (Hons) from Monash University, Melbourne Australia. 
\end{IEEEbiography}

\begin{IEEEbiography}[{\includegraphics[height=1.25in,clip,keepaspectratio]{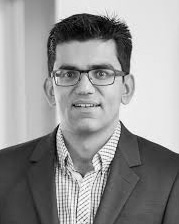}}]{A. Malhotra} (MD, PhD) is a senior neonatologist at Monash Children’s Hospital, and Associate Professor (Research)/ NHMRC Fellow at Monash University, Melbourne, Australia. He has a large research program, with interests in neonatal lung and brain injury, with more than \$7 million in research funding. He has published more than 100 peer reviewed articles, and 4 book chapters to date. Together with Dr Fae Marzbanrad, their team researches digital health technologies to improve neonatal cardiorespiratory monitoring.
\end{IEEEbiography}

\begin{IEEEbiography}[{\includegraphics[height=1.25in,clip,keepaspectratio]{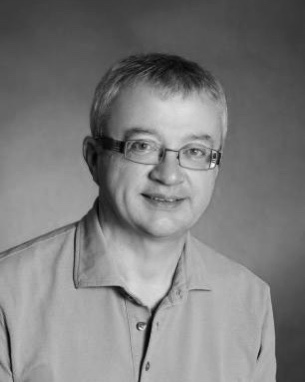}}]{G.A. Dumont} received the Dipl. Ing. degree from Ecole Nationale Supérieure d’Arts et Metiers, Paris, France, in 1973, and the Ph.D. degree in electrical engineering from McGill University, Montreal, QC, Canada, in 1977. He was with Tioxide, France, from 1973 to 1974, and again from 1977 to 1979. He was with Paprican from 1979 to 1989, first in Montreal and then in Vancouver. In 1989, he joined the Department of Electrical and Computer Engineering, University of British Columbia, where he is a Professor and Distinguished University Scholar. From 2000 to 2002, he was the Associate Dean, Research for the Faculty of Applied Science Since 2008 he has been an Associate Member of the UBC Department of Anesthesiology Pharmacology and Therapeutics. He also is a Principal Investigator at the BC Children’s Hospital Research Institute and co-founder and co-Director of the Digital Health Innovation Laboratory (DHIL).
His current research interests include patient monitoring; signal processing for physiological monitoring; physiological closed-loop control systems such as automated drug delivery in anesthesia; circadian rhythms; global and mobile health; non-contact patient vital sign assessment; and brain monitoring via electroencephalography and near-infrared spectrometry.
Dr. Dumont was awarded a 1979 IEEE Transactions on Automatic Control Honorable Paper Award; a 1985 Paprican Presidential Citation; a 1990 UBC Killam Research Prize; the 1995 CPPA Weldon Medal; the 1998 Universal Dynamics Prize for Leadership in Process Control Technology; the IEEE Control Systems Society 1998 Control Systems Technology Award; three NSERC Synergy Awards, the latest one in 2016 for the development of the Phone Oximeter; the 2010 Brockhouse Canada Prize for Interdisciplinary Research in Science and Engineering. In 2011–12, and again in 2018-19, he was a UBC Peter Wall Distinguished Scholar in Residence. In 2020 he was awarded the IEEE Control Systems Society Transition to Practice Award. He has been a Fellow of the IEEE since 1998, and in 2017 he was elected a Fellow of the International Federation of Automatic Control as well as a Fellow of the Royal Society of Canada.

\end{IEEEbiography}

\begin{IEEEbiography}[{\includegraphics[height=1.25in,clip,keepaspectratio]{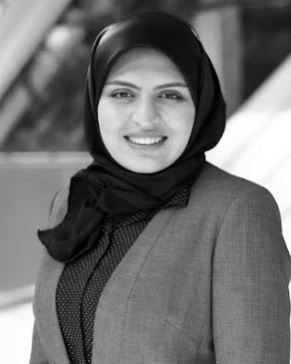}}]{F. Marzbanrad} received her PhD from University of Melbourne, Australia in 2016. She is currently a senior member of IEEE, lecturer and head of Biomedical Signal Processing Lab in the department of electrical and computer systems engineering, Monash University, Australia. Her research interests include biomedical signal processing, machine learning, affordable medical technologies and mobile-health.
\end{IEEEbiography}

\balance 
\vfill

\end{document}